\documentclass[12pt,preprint]{emulateapj}

\begin{document}
\title{A Census of Star-Forming Galaxies in the $z\sim9$-10 Universe
  based on HST+Spitzer Observations Over 19 CLASH clusters: Three
  Candidate $z\sim9$-10 Galaxies and Improved Constraints on the Star
  Formation Rate Density at $z\sim9$\altaffilmark{1}}
\author{R. J. Bouwens\altaffilmark{2,3}, L. Bradley\altaffilmark{4},
A. Zitrin\altaffilmark{5}, D. Coe\altaffilmark{4}, M. Franx\altaffilmark{2},
W. Zheng\altaffilmark{6}, R. Smit\altaffilmark{2}, O. Host\altaffilmark{7},
M. Postman\altaffilmark{4}, L. Moustakas\altaffilmark{8}, 
I. Labb{\'e}\altaffilmark{2}, M. Carrasco\altaffilmark{5,9}, 
A. Molino\altaffilmark{10}, M. Donahue\altaffilmark{11}, 
D.D. Kelson\altaffilmark{12}, M. Meneghetti\altaffilmark{13}, 
N. Ben{\'{\i}}tez\altaffilmark{10}, 
D. Lemze\altaffilmark{6}, K. Umetsu\altaffilmark{14},
T. Broadhurst\altaffilmark{15}, J. Moustakas\altaffilmark{16}, 
P. Rosati\altaffilmark{17}, S. Jouvel\altaffilmark{18}, 
M. Bartelmann\altaffilmark{5}, H. Ford\altaffilmark{6}, 
G. Graves\altaffilmark{19}, C. Grillo\altaffilmark{7}, 
L. Infante\altaffilmark{9}, Y. Jimenez-Teja\altaffilmark{10}, 
O. Lahav\altaffilmark{20}, 
D. Maoz\altaffilmark{21}, E. Medezinski\altaffilmark{6}, 
P. Melchior\altaffilmark{22}, 
J. Merten\altaffilmark{8}, M. Nonino\altaffilmark{23},
S. Ogaz\altaffilmark{4}, 
S. Seitz\altaffilmark{24}}
\altaffiltext{1}{Based on observations
  made with the NASA/ESA Hubble Space Telescope, which is operated by
  the Association of Universities for Research in Astronomy, Inc.,
  under NASA contract NAS 5-26555.}
\altaffiltext{2}{Leiden Observatory, Leiden University}
\altaffiltext{3}{University of California, Santa Cruz}
\altaffiltext{4}{Space Telescope Science Institute}
\altaffiltext{5}{Universitat Heidelberg}
\altaffiltext{6}{The Johns Hopkins University}
\altaffiltext{7}{Dark Cosmology Centre, Niels Bohr Institute, University of Copenhagen}
\altaffiltext{8}{JPL, California Institute of Technology}
\altaffiltext{9}{Universidad Catolica de Chile}
\altaffiltext{10}{Instituto de Astrof\'isica de Andaluc\'ia}
\altaffiltext{11}{Michigan State University}
\altaffiltext{12}{The Carnegie Institute for Science; Carnegie Observatories}
\altaffiltext{13}{INAF, Osservatorio Astronomico di Bologna}
\altaffiltext{14}{Academia Sinica, Institute of Astronomy \& Astrophysics}
\altaffiltext{15}{University of the Basque Country}
\altaffiltext{16}{Siena College}
\altaffiltext{17}{University of Ferrara}
\altaffiltext{18}{Institut de Ciències de l'Espai (IEEC-CSIC)}
\altaffiltext{19}{University of California, Berkeley}
\altaffiltext{20}{University College London}
\altaffiltext{21}{Tel Aviv University}
\altaffiltext{22}{The Ohio State University}
\altaffiltext{23}{INAF, Osservatorio Astronomico di Trieste}
\altaffiltext{24}{Universitas Sternwarte, M\"unchen}
\begin{abstract}
We utilise a two-color Lyman-Break selection criterion to search
for $z\sim9$-10 galaxies over the first 19 clusters in the CLASH
program.  A systematic search yields three $z\sim9$-10
candidates.  While we have already reported the most robust of
these candidates, MACS1149-JD, two additional $z\sim9$ candidates
are also found and have $H_{160}$-band magnitudes of
$\sim$26.2-26.9.  A careful assessment of various sources of
contamination suggests $\lesssim$1 contaminants for our
$z\sim9$-10 selection.  To determine the implications of these
search results for the LF and SFR density at $z\sim9$, we
introduce a new differential approach to deriving these
quantities in lensing fields.  Our procedure is to derive the
evolution by comparing the number of $z\sim9$-10 galaxy
candidates found in CLASH with the number of galaxies in a
slightly lower redshift sample (after correcting for the
differences in selection volumes), here taken to be $z\sim8$.
This procedure takes advantage of the fact that the relative
volumes available for the $z\sim8$ and $z\sim9$-10 selections
behind lensing clusters are not greatly dependent on the details
of the lensing models.  We find that the normalization of the UV
LF at $z\sim9$ is just $0.28_{-0.20}^{+0.39}\times$ that at
$z\sim8$, $\sim$1.4$_{-0.8}^{+3.0}$$\times$ lower than
extrapolating $z\sim4$-8 LF results.  While consistent with the
evolution in the $UV$ LF seen at $z\sim4$-8, these results
marginally favor a more rapid evolution at $z>8$.  Compared to
similar evolutionary findings from the HUDF, our result is less
insensitive to large-scale structure uncertainties, given our
many independent sightlines on the high-redshift universe.
\end{abstract}
\keywords{galaxies: evolution --- galaxies: high-redshift}

\section{Introduction}

Since the discovery of large numbers of $z\sim3$ galaxies with the
Lyman-break selection technique 17 years ago (Steidel et al.\ 1996),
there has been a persistent effort to use the latest facilities to
identify galaxies at higher and higher redshifts through photometric
selections and follow-up spectroscopy.  These efforts allow us to
probe galaxies during the epoch of reionization to ascertain what role
they may have in driving this process.  Progressively, the
high-redshift frontier has been extended to $z\sim4$-5 (e.g., Madau et
al.\ 1996; Steidel et al.\ 1999), $z\sim6$ (e.g., Stanway et
al.\ 2003; Bouwens et al.\ 2003; Dickinson et al.\ 2004), $z\sim7$
(e.g., Bouwens et al.\ 2004; Yan \& Windhorst 2004; Bouwens \&
Illingworth 2006b; Iye et al.\ 2006; Fontana et al.\ 2009; Schenker et
al.\ 2012), and $z\sim8$ (e.g., Bouwens et al.\ 2010; McLure et
al.\ 2010; Bunker et al.\ 2010; Yan et al.\ 2010).

The current frontier for identifying high-redshift galaxies now seems
to lie firmly at $z\sim10$, with three distinct $z\sim10$ galaxy
candidates having been reported.\footnote{Following the initial
  submission of this paper, seven additional $z\sim10$ candidates have
  been identified: a $z\sim 9.5$ candidate from Ellis et al.\ (2013)
  over the HUDF, a $z\sim9.8$ candidate from Oesch et al.\ (2014) over
  the HUDF, four $z\sim9.5$-10.2 candidates from Oesch et al.\ (2014)
  over CANDELS, and a triply-lensed $z\sim9.8$ candidate from Zitrin
  et al.\ (2014) behind Abell 2744.}  Bouwens et al.\ (2011a)
presented the discovery of a plausible $z\sim10.3$ galaxy in the full
two-year HUDF09 observations over the HUDF (see also Oesch et
al.\ 2012a).  More recently, Zheng et al.\ (2012: hereinafter Z12)
presented evidence for a highly-magnified $z\sim9.6$ galaxy within the
524-orbit CLASH program (Postman et al.\ 2011), and Coe et al.\ (2013:
hereinafter C13) reported the discovery of an even higher redshift
triply-lensed $z\sim10.8$ galaxy.

Despite the very interesting nature of earlier exploratory work, the
total number of $z\sim9$-11 galaxies is small, and hence it is still
somewhat challenging to obtain accurate constraints on how rapidly the
luminosity function (LF) or star formation rate (SFR) density evolved
in the very early universe, at $z>8$.  Earlier $z\sim10$ searches
using the very deep HUDF09 data (Bouwens et al.\ 2011a; Oesch et
al.\ 2012a) found tentative evidence for a deficit of $z\sim10$
galaxies relative to simple extrapolations from lower redshifts,
pointing towards a very rapid evolution in the UV LF and SFR density
at $z>8$ (Oesch et al.\ 2012a).  A rapid evolution of the $UV$ LF at
$z>8$ is supported by several theoretical models (Trenti et al.\ 2010;
Lacey et al.\ 2011), but may be in some tension with the discovery of
one bright, multiply-lensed $z\sim10.8$ galaxy in the CLASH program
(C13), since one might have expected such sources to be quite rare
assuming a rapid evolution of the $UV$ LF.

Fortunately, there is an ever increasing quantity of observations now
available to identify $z\sim9$-10 galaxies.  One noteworthy near-term
opportunity exists in the moderately deeper WFC3/IR observations
acquired over the HUDF (GO 12498: Ellis et al.\ 2013).  This program
has made it possible to extend $z\sim9$-10 samples in the HUDF deeper
by $\sim$0.4 mag while increasing the number of sources by a factor of
2-4 (McLure et al.\ 2013).  However, another significant opportunity
exists in ongoing observations over lensing clusters, as part of the
524-orbit CLASH program (Postman et al.\ 2012).  The initial discovery
papers of Z12 and C13 only reported on the brightest and most robust
$z\sim10$ and $z\sim11$ galaxy candidates from the CLASH program;
however, it should be possible to extend these searches somewhat
fainter by $\sim$0.5-1.0 mag to the magnitude limit of the survey
($\sim$27 AB mag).  At such magnitudes, we would expect to identify
other plausible $z\sim9$-10 galaxies, potentially increasing the
overall sample size to $\sim$3-5 sources in total.

The purpose of this paper is to capitalize on the opportunity that
exist within lensing clusters from the CLASH program.  A deeper search
for $z\sim9$-10 galaxies can be performed in a reasonably reliable
manner taking full advantage of the substantial observations with
Spitzer/IRAC instrument over the CLASH program (Egami et al.\ 2008;
Bouwens et al.\ 2011c), allowing us to distinguish potential
star-forming galaxy candidates at $z\sim9$-10 from lower-redshift
interlopers.  We also incorporate HST observations over 2 more
clusters from the CLASH program (utilizing a total of 19 clusters) to
expand the total search area by 50\% and 10\% over what was considered
in Z12 and C13, respectively.

The plan for this paper is as follows.  In \S2, we describe our
observational data set.  In \S3, we discuss our procedure for catalog
creation, the selection of $z\sim9$-10 galaxy candidates, quantifying
their properties, and estimating the extent to which contamination may
be a concern for our selection.  In \S4, we introduce a new
differential approach to derive the evolution in the $UV$ LF and SFR
density at $z\gtrsim9$ and then apply it to our search results at
$z\sim9$.  Finally, in \S5, we summarize the results from this paper
and offer a prospective.  Throughout this work, we quote results in
terms of the luminosity $L_{z=3}^{*}$ Steidel et al.\ (1999) derived
at $z\sim3$: $M_{1700,AB}=-21.07$.  We refer to the HST F225W, F390W,
F435W, F475W, F606W, F625W, F775W, F814W, F850LP, F105W, F110W, F125W,
F140W, and F160W bands as $UV_{225}$, $U_{390}$, $B_{435}$, $g_{475}$,
$V_{606}$, $r_{625}$, $i_{775}$, $I_{814}$, $z_{850}$, $Y_{105}$,
$J_{110}$, $J_{125}$, $JH_{140}$, and $H_{160}$, respectively, for
simplicity.  Where necessary, we assume $\Omega_0 = 0.3$,
$\Omega_{\Lambda} = 0.7$, $H_0 = 70\,\textrm{km/s/Mpc}$.  All
magnitudes are in the AB system (Oke \& Gunn 1983).

\section{Observational Data}

Our primary dataset for this study are the 20-orbit HST observations
over the first 19 clusters with data from the 524-orbit CLASH
multi-cycle treasury program (Postman et al.\ 2012: see
Table~\ref{tab:obsdata}).  The HST observations over each of the CLASH
clusters is typically distributed over 16 different bands using the
WFC3/UVIS camera, the Advanced Camera for Surveys (ACS) wide field
camera, and the WFC3/IR camera.  These observations extend from
0.2$\mu$m ($UV_{225}$) to 1.6$\mu$m ($H_{160}$) and reach to depths to
26.4-27.7 AB mag ($5\sigma$: 0.4$''$-diameter aperture) depending upon
the passband.

Our reductions of these data were conducted using standard procedures,
aligned, and then drizzled on the same frame (0.065$''$ pixel scale)
with the multidrizzle software (Koekemoer et al.\ 2003).  The FWHM for
the PSF is $\sim$0.1$''$ in the WFC3/UVIS or ACS observations and
$\sim$0.16-0.17$''$ for the WFC3/IR observations.

The typical area available over each cluster to search for $z\sim9$-10
galaxies is $\sim$4 arcmin$^2$ and is dictated by the area available
within the WFC3/IR field-of-view.  In total, we make use of $\sim$77
arcmin$^2$ over the first 19 CLASH clusters to search for $z\sim9$-10
galaxies.  This corresponds to an approximate search volume of
$\sim$7000 Mpc$^{3}$ (comoving) at $z\sim9$ to probe faint, highly
magnified $\mu>5$ galaxies (assuming $\sim$25\% of our WFC3/IR area is
high magnification $\mu\gtrsim5$ and a $\Delta z\sim1$ width for our
redshift selection window: see Figure~\ref{fig:zdist}).  To ensure
that we have the maximum depth and filter coverage available for
candidates uncovered in our search, we do not consider the small
amount of data over each cluster with observations in only one of the
two roll angles used for the CLASH program (see figure 11 of Postman
et al.\ 2012 for an illustration of the two roll-angle strategy).

\begin{deluxetable}{ccc}
\tablewidth{0cm}
\tabletypesize{\footnotesize}
\tablecaption{The 19 cluster fields from the CLASH program considered in the present $z\sim9$ search.\label{tab:obsdata}}
\tablehead{\colhead{Cluster} & \colhead{Redshift} & \colhead{High Magnification\tablenotemark{a}}}
\startdata
Abell 209 & 0.206 & \\
Abell 383 & 0.187 & \\
Abell 611 & 0.288 & \\
Abell 2261 & 0.224 & \\
MACS0329.7$-$0211 & 0.450 \\
MACS0416.1$-$2403 & 0.42 & Y\\
MACS0647.8+7015 & 0.584 & Y\\
MACS0717.5+3745 & 0.548 & Y\\ 
MACS0744.9+3927 & 0.686 \\
MACS1115.9+0129 & 0.352 \\
MACS1149.6+2223 & 0.544 & Y\\
MACS1206.2$-$0847 & 0.440 \\
MACSJ1720.3+3536 & 0.391 \\
MACSJ1931.8$-$2635 & 0.352 \\
MACSJ2129.4$-$0741 & 0.570 & Y\\
MS2137$-$2353 & 0.313 \\
RXJ1347.5$-$1145 & 0.451 \\
RXJ1532.9+3021 & 0.345 \\
RXJ2129.7+0005 & 0.234 \\
\enddata
\tablenotetext{a}{Clusters in the CLASH program were selected based on
  either their x-ray or magnification properties (Postman et
  al.\ 2012).  Clusters marked here with a ``Y'' were included because
  of their magnification properties.}
\end{deluxetable}

Each of the CLASH clusters also has a substantial amount of
observations with the Spitzer/IRAC instrument (Fazio et al.\ 2004).
The typical integration times range from $\sim$3.5 hours per IRAC band
from the ICLASH program (GO \#80168: Bouwens et al.\ 2011c) to $\sim$5
hours per IRAC band from the Spitzer IRAC Lensing Survey program (GO
\#60034: PI Egami).  Even deeper observations are available over from
the Surfs'Up program (Brada{\v c} et al.\ 2014: 30 hours), Frontier
Field program (T. Soifer and P. Capak: 50 hours), and follow-up
observations on MACS0647 and MACS1720 (PI Bouwens [90213]: 11/24
hours; Coe [10140]: 56 hours).  These observations reach to 1$\sigma$
depths of $\sim$26.2-27.4 mag in both the 3.6$\mu$m and 4.5$\mu$m IRAC
channels, allowing us to set useful constraints on the color of
possible $z\sim9$-10 candidates redward of the break.  The FWHM for
the IRAC PSF at 3.6$\mu$m and 4.5$\mu$m is $\sim$1.8$''$.  We reduced
the Spitzer/IRAC observations using the public MOPEX software
available from the Spitzer Science Center (Makovoz et al.\ 2005),
excluding roll angles where artifacts from bright stars had an impact
on the photometry of the candidate $z\sim9$-10 galaxies under study.
The reductions were drizzled onto a common output frame ($0.6''$-pixel
scale).

\section{Results}

\subsection{Catalog Construction}

Our procedure for constructing catalogs is similar to that previously
utilized by Bouwens et al.\ (2007, 2011b, 2012b).  These catalogs are
distinct from those distributed as part of the CLASH program, but
overall the results are in very good agreement.

We provide a brief outline of the procedure we use here.  More details
are provided in several of our previous publications (e.g., Bouwens et
al.\ 2007, 2011b, 2012b). SExtractor (Bertin \& Arnouts 1996) is run
in dual-image mode, using the square root of the $\chi^2$ image
(Szalay et al.\ 1999) to detect sources and the PSF-matched images for
photometry.  The $\chi^2$ image (similar to a coadded frame) is
constructed from the imaging observations in the two passbands where
we expect $z\sim9$ candidates to show significant signal, i.e., the
$JH_{140}$ and $H_{160}$ bands.  For the photometry, PSF-matching is
done to the WFC3/IR $H_{160}$-band.  Fluxes and colors of sources are
measured in apertures that scale with the size of sources, as
recommended by Kron (1980) and using a Kron factor of 1.2.  The
small-aperture fluxes are then corrected to total magnitudes in two
steps.  First the excess flux around the source in a larger scalable
aperture (Kron factor 2.5) is derived based on the square root of
$\chi^2$ image and this correction is applied to the measured fluxes
in all HST bands.  Second, a correction is made for the expected light
outside the larger scaled aperture and on the wings of the PSF using
the tabulated encircled energy distribution (e.g., from Sirianni et
al.\ 2005).

The measurement of IRAC fluxes is important for a more secure
identification of $z\sim9$ candidates in our fields, since it allows
us to quantify the approximate spectral slope of the sources redward
of the spectral break observed at $\sim$1.2$\mu$m and therefore
distinguish potential star-forming galaxies at $z\sim9$-10 from
interlopers at $z\sim1$-2.  IRAC photometry can be challenging due to
the significant overlap between nearby sources in existing data.
Fortunately, there are well-established procedures to use the
positions and spatial profiles of sources in available HST
observations to model the IRAC image observations and extract fluxes
(e.g., Shapley et al.\ 2005; Labb{\'e} et al.\ 2006; Grazian et al.\ 2006;
Laidler et al.\ 2007).

Here we make use of the \textsc{Mophongo} software (Labb{\'e} et
al.\ 2006, 2010a, 2010b, 2013) to do photometry on sources in our
fields, given the confusion.  Since this software has been presented
more extensively in other places, we only include a brief description
here.  The most important step for doing photometry on faint sources
with this software is to remove confusion from neighboring sources.
This is accomplished by using the deep WFC3/IR observations as a
template to model the positions and isolated flux profiles of the
foreground sources.  These flux profiles are then convolved to match
the IRAC PSF and then simultaneously fit to the IRAC imaging data
leaving only the fluxes of the sources as unknowns.  The best-fit
model is then used to subtract the flux from neighboring sources and
normal aperture photometry is performed on sources in a
2.5$''$-diameter aperture.  The measured $3.6\mu$m and $4.6\mu$m
fluxes are then corrected to account for the light on the wings of the
IRAC PSF (typically the correction is a factor of $\sim$2.2).

\begin{figure}
\epsscale{1.15} \plotone{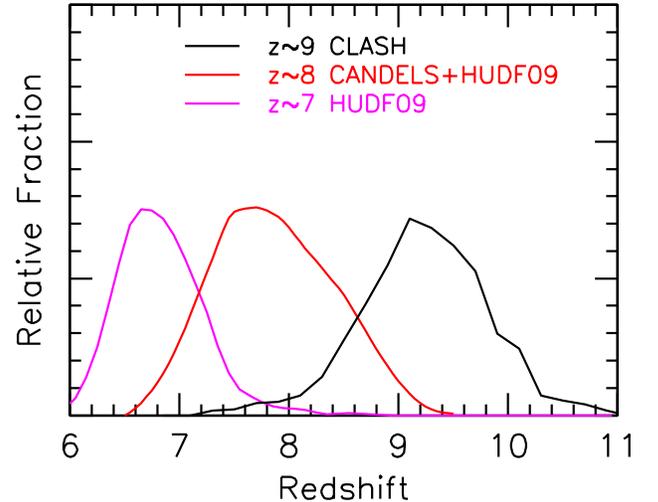}
\caption{The redshift distribution we would expect for our present
  $z\sim9$ selection based on the simulations we run in \S4.3.  These
  simulations allow us to assess the relative selection volume for our
  $z\sim9$ selection and our comparison sample at $z\sim8$.  The mean
  redshift for our selection is 9.2.  Our $z\sim9$ selection cuts off
  at $z>10$ due to our use of a $JH_{140}-H_{160}<0.5$ criterion
  (\S3.2: see also Figure~\ref{fig:selcrit}).  For context, we also
  show the expected redshift distributions for the $z\sim7$ and
  $z\sim8$ selections of Bouwens et al.\ (2011b) and Oesch et
  al.\ (2012b), respectively.\label{fig:zdist}}
\end{figure}

\begin{figure*}
\epsscale{1.15}
\plotone{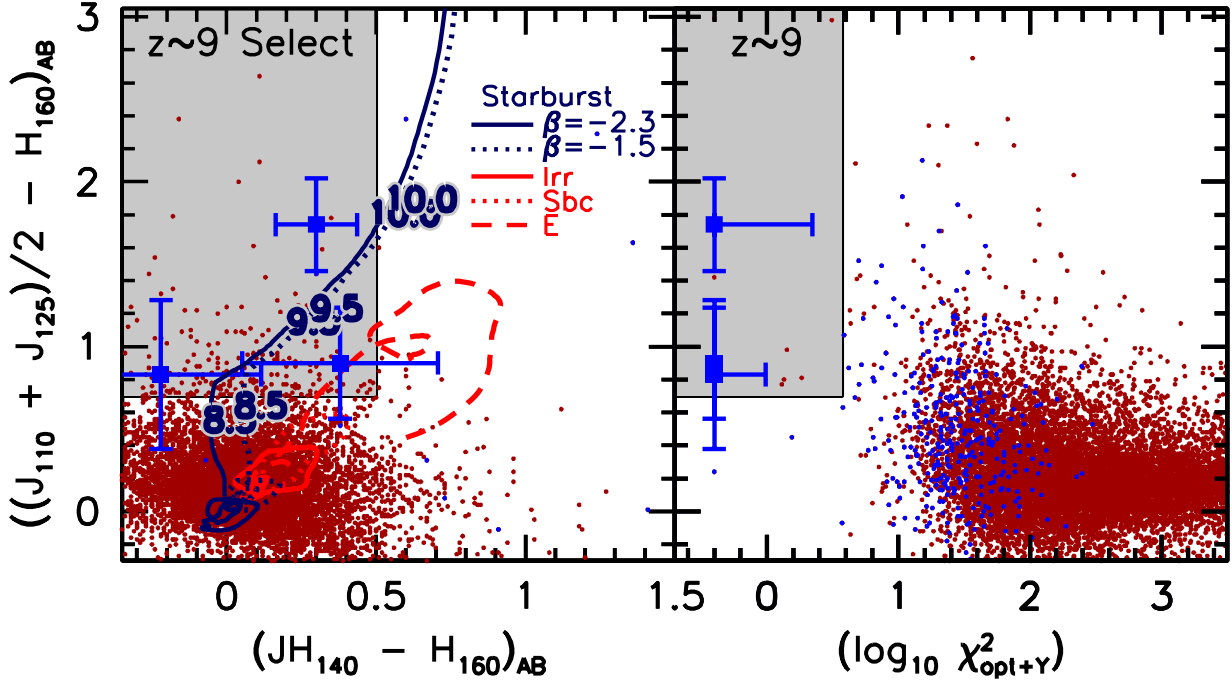}
\caption{Selection criteria used here to identify $z\sim9$-10 galaxies
  over the CLASH program.  (\textit{left}) The $((J_{110}+J_{125})/2 -
  H_{160})_{AB}$ vs. $(JH_{140} - H_{160})_{AB}$ diagram shows the
  first of our two primary criteria we use to identify $z\sim9$-10
  galaxies from the CLASH program.  Selected sources must fall in the
  gray region defined by two LBG-like color criteria, with a
  $(J_{110}+J_{125})/2 - H_{160}>0.7$ criterion defining the Lyman
  break and a $JH_{140}-H_{160}<0.5$ criterion providing a constraint
  on the spectral slope redward of the break.  The large blue squares
  show the sources that made it into our $z\sim9$-10 sample.  The
  error bars on these points are the $1\sigma$ uncertainties.  The
  blue lines show the expected colors for star-forming galaxies with
  varying $UV$-continuum slopes as a function of redshift while the
  red lines show the expected colors for different SED templates at
  lower redshift (Coleman et al.\ 1980).  The small dark red points
  show the colors of sources in our photometric sample where the
  $\chi_{opt+Y}^2$ statistic is $>3.8$.  The blue points show these
  colors for sources where the $\chi_{opt+Y}^2$ statistic is $<3.8$.
  See \S3.2 (and Bouwens et al.\ 2011b) for a definition of the
  $\chi_{opt+Y}^2$ statistic, but it roughly includes a stack of all
  the flux information in the $Y_{105}$ band and bluer bands.
  (\textit{right}) The $((J_{110}+J_{125})/2 - H_{160})_{AB}$
  vs. $\chi_{opt+Y}^2$ diagram shows the second of our two primary
  criteria we use to identify $z\sim9$-10 galaxies from the CLASH
  program.  The selected sources must fall in the gray region and
  therefore must show no flux in the optical or $Y_{105}$ bands (i.e.,
  $\chi_{opt+Y}^2<3.8$).  The three selected $z\sim9$ candidates are
  the blue squares.  The dark red points indicate sources in our
  photometric sample which are either detected in the $Y_{105}$ band
  ($>$2$\sigma$) or where the $JH_{140}-H_{160}$ colors are greater
  than 0.5.  The blue points are those sources where neither condition
  is satisfied.  This figure is similar to Figure 2 of Oesch et
  al.\ (2012b).  Using both the two-color criteria and our
  $\chi_{opt+Y}^2$ criteria, we observe a clear separation between our
  $z\sim9$-10 candidates and the bulk of our photometric sample.
  While we cannot completely rule out certain classes of
  lower-redshift galaxies contaminating our selection (note that the
  color-color track for early-type galaxies overlaps our selection
  window in the left panel), the volume density of such contaminants
  would seem to be lower than that of bright $z\sim9$-10 galaxies (see
  Appendix A).\label{fig:selcrit}}
\end{figure*}

\subsection{Source Selection}

In this paper, we adopt a two-color Lyman-break selection to search
for promising $z\sim9$-10 galaxy candidates in the CLASH program.
This work takes advantage of the sharp break in the spectrum of
star-forming galaxies due to absorption by neutral hydrogen.  Many
years of spectroscopic work have demonstrated that the Lyman-break
selection technique provides us with a very efficient means of
identifying high-redshift galaxies (Steidel et al.\ 1996; Steidel et
al.\ 2003; Bunker et al.\ 2003; Dow-Hygelund et al.\ 2007; Popesso et
al.\ 2009; Vanzella et al.\ 2009; Stark et al.\ 2010), with generally
minimal contamination, albeit with a few notable exceptions at
brighter magnitudes (e.g., Steidel et al.\ 2003; Bowler et al.\ 2012;
Hayes et al.\ 2012).  In the latter case, deeper mid-IR data can be
valuable for guarding against such contamination.

In analogy with lower-redshift Lyman-break selections (e.g.,
Giavalisco et al.\ 2004; Bouwens et al.\ 2007; Bouwens et al.\ 2011b),
we devised the following two-color $z\sim9$-10 selection for the CLASH
cluster fields:
\begin{displaymath}
((J_{110}+J_{125})/2 - H_{160} > 0.7)\wedge(JH_{140}-H_{160} < 0.5)
\end{displaymath}
where $\wedge$ represents the logical \textbf{AND} symbol.  This
criterion is very similar to the criteria previously presented in Z12,
i.e., $(J_{110}-JH_{140} > 0.5) \wedge(JH_{140}- H_{160}<0.5)$, but
probe to slightly higher redshift sources on average, also folding in
information from the redder $J_{125}$-band filter and requiring a
sharper break in the spectrum.  In general, it makes sense to combine
the flux information from both the $J_{110}$ and $J_{125}$ bands to
search for $z\gtrsim9$ candidates because of their similar red-side
cut-offs at $1.4\mu$m.  In applying the above criteria, the magnitudes
of sources not detected at $1\sigma$ are set to their $1\sigma$ upper
limits.

\begin{figure*}
\epsscale{1.15}
\plotone{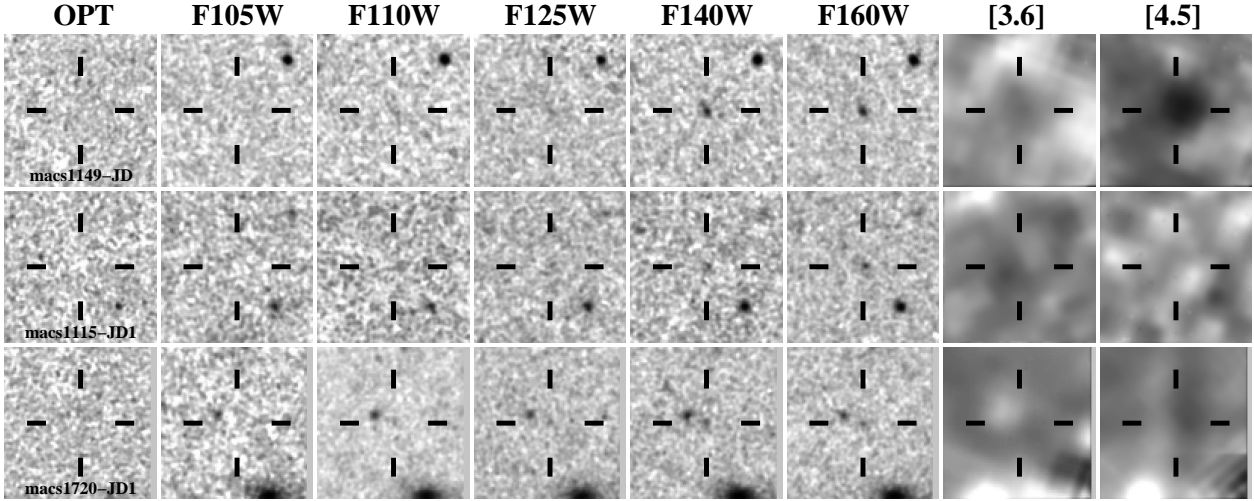}
\caption{Postage stamp images ($6.6''\times6.6''$) of the three
  $z\sim9$ galaxy candidates we identify in the current 19-cluster
  CLASH observations.  The source in uppermost row is the same
  $z\sim9.6$ candidate as we reported in Z12 (though our redshift
  estimate for this source is a very consistent $z\sim9.7$: see
  \S3.4).  The leftmost postage stamp shows a stack of the deep ACS
  $B_{435}+g_{475}+V_{606}+r_{625}+i_{775}+I_{814}+z_{850}$ optical
  observations, while the other stamps show the observations in
  specific HST WFC3/IR and Spitzer/IRAC bands.  On the IRAC postage
  stamps, flux from neighboring sources (as derived by
  \textsc{Mophongo}) has been subtracted off.  All three of our
  $z\sim9$ candidates are detected at $>6.8\sigma$ in a coadded
  $JH_{140}+H_{160}$ image (0.35$''$-diameter aperture: see
  Table~\ref{tab:properties}).  The Spitzer fluxes we measure for the
  sources are sufficiently faint, as to substantially prefer a $z>6$
  solution for the sources rather than a low redshift solution.  None
  of the sources show any significant detections in the optical ACS
  observations.\label{fig:stamps}}
\end{figure*}

\begin{deluxetable*}{ccccc}
\tablewidth{13cm}
\tabletypesize{\footnotesize}
\tablecaption{Coordinates, Estimated Redshifts and Magnification Factors, and Photometry for Present $z\sim9$ Sample.\tablenotemark{a}\label{tab:properties}}
\tablehead{\colhead{} & \colhead{MACS1149-JD\tablenotemark{b}} & \colhead{MACSJ1115-JD1} & \colhead{MACSJ1720-JD1} & \colhead{Stack\tablenotemark{c}}}
\startdata
R.A. & 11:49:33.58 & 11:15:54.50 & 17:20:12.76 & --- \\
Decl & 22:24:45.7 & 01:29:47.9 & 35:36:17.5 & --- \\
$z_{photo}$\tablenotemark{d} & $9.7_{-0.1}^{+0.1}$\tablenotemark{e} & $9.2_{-0.8}^{+0.4}$ & $8.9_{-0.5}^{+0.3}$ & --- \\
Magnification & 14.5$_{-1.0}^{+4.2}$ & 9.3$_{-3.6}^{+5.8}$ & 5.0$_{-0.7}^{+4.7}$ & --- \\
S/N ($JH_{140}+H_{160}$)\tablenotemark{f} & 15.4 & 7.8 & 6.9 & --- \\
$U_{390}$ & $-$8$\pm$25 & $-$14$\pm$37 & 16$\pm$32 & 1$\pm$18 \\
$B_{435}$ & $-$1$\pm$26 & $-$117$\pm$39 & 4$\pm$32 & $-$35$\pm$19 \\
$g_{475}$ & $-$3$\pm$19 & $-$23$\pm$25 & $-$10$\pm$20 & $-$12$\pm$12 \\
$V_{606}$ & $-$9$\pm$14 & $-$0$\pm$35 & $-$11$\pm$28 & $-$7$\pm$16 \\
$r_{625}$ & $-$27$\pm$22 & 10$\pm$24 & $-$9$\pm$17 & $-$6$\pm$11 \\
$i_{775}$ & 0$\pm$27 & 49$\pm$47 & $-$35$\pm$38 & 0$\pm$22 \\
$I_{814}$ & $-$3$\pm$11 & $-$13$\pm$20 & $-$27$\pm$17 & $-$16$\pm$10 \\
$z_{850}$ & $-$38$\pm$34 & $-$32$\pm$55 & 4$\pm$39 & $-$15$\pm$25 \\
$Y_{105}$ & $-$3$\pm$17 & $-$39$\pm$23 & $-$20$\pm$20 & $-$21$\pm$12 \\
$J_{110}$ & 27$\pm$13 & 37$\pm$19 & 22$\pm$13 & 26$\pm$8 \\
$J_{125}$ & 56$\pm$16 & 63$\pm$21 & 44$\pm$16 & 49$\pm$10 \\
$JH_{140}$ & 146$\pm$15 & 80$\pm$22 & 80$\pm$15 & 86$\pm$10 \\
 & (=26.0$\pm$0.1) & (=26.6$\pm$0.3) & (=26.6$\pm$0.2) & (=26.6$\pm$0.1) \\
$H_{160}$ & 193$\pm$15 & 115$\pm$19 & 66$\pm$16 & 100$\pm$10 \\
 & (=25.7$\pm$0.1) & (=26.2$\pm$0.2) & (=26.9$\pm$0.3) & (=26.4$\pm$0.1) \\
$[3.6]$ & 164$\pm$41\tablenotemark{g} & 356$\pm$110 & $-$39$\pm$123 & 160$\pm$56 \\
$[4.5]$ & 342$\pm$66\tablenotemark{g} & $-$52$\pm$114 & 195$\pm$124 & 161$\pm$60 \\
$\frac{1}{2}([3.6]+[4.5])$\tablenotemark{h} & 253$\pm$38 & 152$\pm$79 & 78$\pm$87 & 161$\pm$41 \\
 & (=25.4$\pm$0.2) & (=26.0$\pm$0.5) & ($>$26.7) & (=25.9$\pm$0.3) 
\enddata
\tablenotetext{a}{The fluxes in this table are in units of nJy.}
\tablenotetext{b}{The same candidate as is presented in Z12.  The fluxes presented in this table were derived independently from those presented in Z12, but are very similar in general.}
\tablenotetext{c}{This column gives the average fluxes in all HST+IRAC
  bands blueward of $0.4\mu$m for the three $z\sim9$ candidates in our
  selection.  The fluxes of each source are rescaled such that its
  average $JH_{140}$+$H_{160}$ flux matches the average
  $JH_{140}+H_{160}$ flux of the sample (prior to rescaling).}
\tablenotetext{d}{These photometric redshift estimates are based on
  the EAZY photometric redshift software (Brammer et al.\ 2008: see
  \S3.4).  In \S3.4, we also provide photometric redshift estimates
  for sources using BPZ and Le PHARE.}
\tablenotetext{e}{Z12 prefer a slightly lower
  redshift of 9.6 for this source based on the photometry, but within
  the uncertainties, the present estimate is fully consistent with
  that given in Z12.}
\tablenotetext{f}{S/N of our $z\sim9$ candidates in the $JH_{140}$ and
  $H_{160}$ bands added in quadrature (0.35$''$-diameter circular
  aperture).  The S/N limit for our $z\sim9$ selection was 6.0.  Our
  highest S/N candidates are much less likely to correspond to
  lower-redshift contaminants (see \S3.5, Figure~\ref{fig:seds}, and
  Figure~\ref{fig:surfdens}).}
\tablenotetext{g}{The fluxes we measure in the Spitzer/IRAC $3.6\mu$m
  and $4.5\mu$m channels are very similar to that measured by Brada{\v
    c} et al.\ (2014), i.e., $25.70\pm 0.17 \pm 0.49$ (196$\pm$32 nJy)
  in $3.6\mu$m channel and $25.01\pm 0.08 \pm 0.21$ (370$\pm$30 nJy)
  in the $4.5\mu$m channel.}
\tablenotetext{h}{We also presented an average of the
  Spitzer/IRAC $3.6\mu$m and $4.5\mu$m fluxes due to the limited S/N
  of each of these measurements.}
\end{deluxetable*}

It is also important we detect sources at sufficient S/N that we can
rely on the color information (and optical non-detections) to provide
reliable redshift information on the sources and guarantee they are
real.  After some experimentation and extensive simulations (\S3.5),
we elected to require sources in our $z\sim9$ selection be detected at
$\geq$6$\sigma$ in a combined $JH_{140}$ and $H_{160}$ bands (using a
fixed 0.35$''$-diameter aperture).  For significance thresholds less
than $6\sigma$, our simulations (\S3.5) suggest that our $z\sim9$
selection would be subject to significant contamination from lower
redshift interlopers.

To ensure that sources really have no flux in the spectrum blueward of
the Lyman break, we also require sources be undetected
($<$2.5$\sigma$) in the $Y_{105}$ band and any passband blueward of
this.\footnote{Since we combine the optical flux measurements into
  several $\chi^2$ statistics that we use to test the plausibility of
  specific sources as $z\sim9$ candidates, we only adopt a weaker
  2.5$\sigma$ threshold here to avoid unnecessarily excluding many
  plausible $z\sim9$ candidates.}  Moreover, we combine the flux in
all the bluer bands ($U_{390}$, $B_{435}$, $g_{475}$, $V_{606}$,
$r_{625}$, $i_{775}$, $I_{814}$, $z_{850}$, and $Y_{105}$) to
construct a $\chi^2$ statistic for sources in our catalogs and exclude
sources from our selection if the $\chi_{opt+Y}^2$ statistic is
greater than 3.8.  The particular threshold for $\chi_{opt+Y}^2$,
i.e., 3.8, was chosen to keep contamination in our $z\sim9$ sample
relatively low while not overly impacting the completeness of our
samples (see figure 19 from Bouwens et al.\ 2011b for an illustration
of how such a choice can be made).  This criterion is very effective
at guarding against contamination from sources which are consistently
faint in all optical bands; it ensures that sources are not
consistently detected at $>1\sigma$ in more than three optical bands.

Here $\chi^2$ is calculated as follows: $\chi_{opt+Y} ^2 = \Sigma_{i}
\textrm{SGN}(f_{i}) (f_{i}/\sigma_{i})^2$ where $f_{i}$ is the flux in
band $i$ in a consistent aperture, $\sigma_i$ is the uncertainty in
this flux, and SGN($f_{i}$) is equal to 1 if $f_{i}>0$ and $-1$ if
$f_{i}<0$ (Bouwens et al.\ 2011b).  As in Bouwens et al.\ (2011b), we
calculate this $\chi^2$ statistic in three different apertures
(scalable Kron apertures [Kron factor of 1.2], 0.35$''$-diameter
circular apertures, 0.18$''$-diameter circular apertures) to ensure
that there is absolutely no evidence for a significant excess of light
blueward of the break, whether this light be tightly concentrated on
the source itself or more diffuse.  When computing the $\chi^2$
statistic with 0.18$''$-diameter apertures, we use the original
unsmoothed ACS or WFC3/IR images (i.e., before PSF-matching to the
WFC3/IR $H_{160}$-band data) to retain the maximum signal-to-noise for
the purposes of rejecting low-redshift interlopers.

As one final step to ensure that our $z\sim9$ candidates show no
evidence for flux blueward of the break, we construct a second
$\chi^2$ statistic for each source, utilizing only the information in
the three bands immediately blueward of the break, i.e., the
$I_{814}$, $z_{850}$, and $Y_{105}$ bands.  We then exclude any source
which has an $\chi_{I+z+Y}^2$ value greater than 3.  This criterion
provides us with better discrimination against dusty lower-redshift
interlopers (which we would not expect to be detected in the bluer
bands) and serves as an effective complement to our other $\chi^2$
criterion (which is better at discriminating against sources which are
consistently faint in all optical bands).  Sources detected at
$>$2$\sigma$ in the $Y_{105}$-band are also excluded to minimize the
contribution of $z\sim8$ galaxies to our selection.

In Figure~\ref{fig:zdist}, we show the approximate redshift selection
window for our current selection.  Details on how it is calculated
will be presented in \S4.3, but approximately involve adding
artificial sources to the real data with realistic colors, sizes, and
magnitudes, and then attempting to reselect them with the criteria
given above.  The mean redshift we derive for our $z\sim9$-10
selection from the simulations is 9.2.  For context, we also present
the redshift selection windows for samples at $z\sim7$ and $z\sim8$,
as selected by Bouwens et al.\ (2011b) and Oesch et al.\ (2012b),
respectively.

\begin{figure*}
\epsscale{1.15}
\plotone{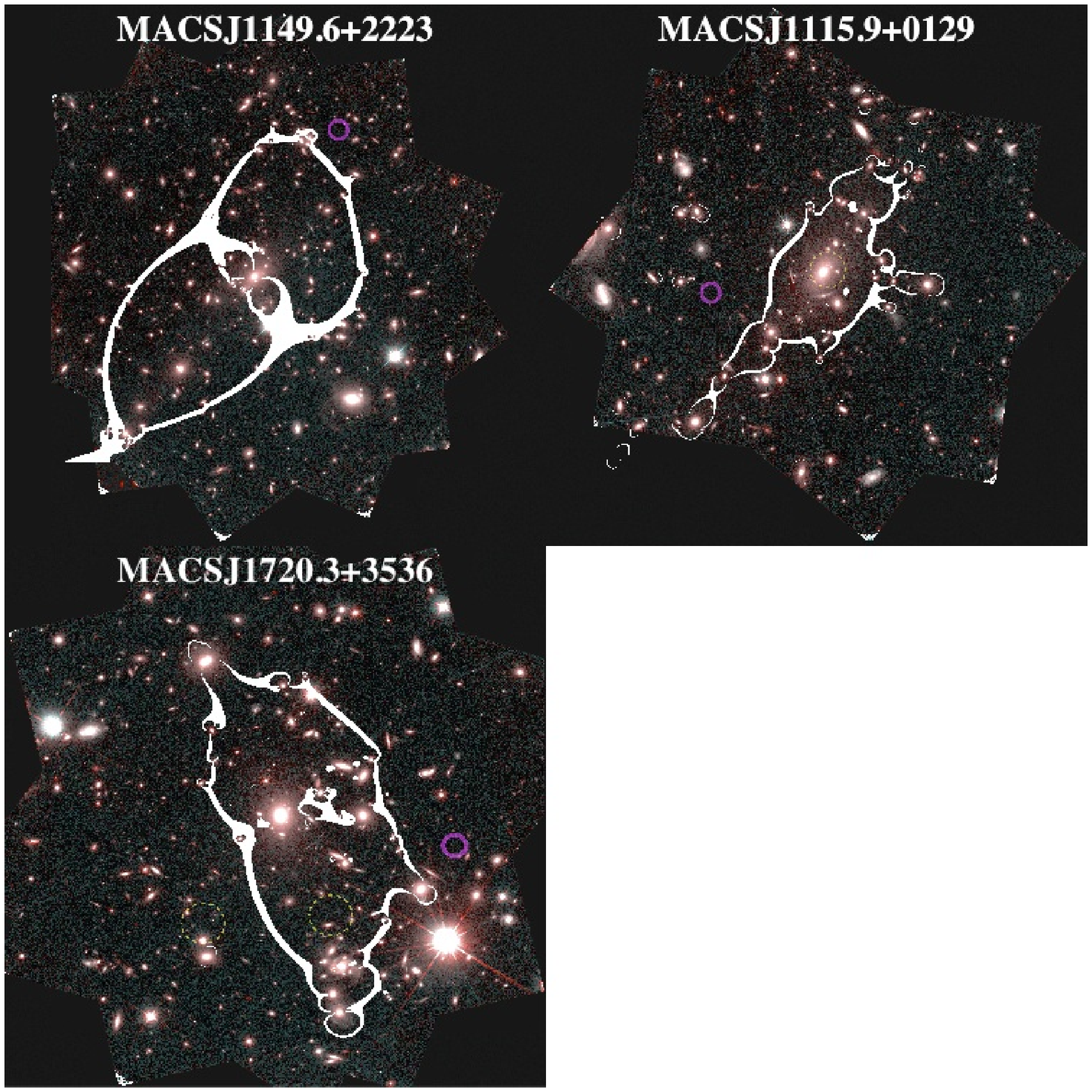}
\caption{Position of the three $z\sim9$ galaxy candidates we identify
  over MACSJ1149.6+2223, MACSJ1115.9+0129, and MACSJ1720.3+3536.  The
  color images shown are based on the HST $I_{814}$ + $H_{160}$
  observations of these clusters with CLASH and are shown over those
  regions with deep WFC3/IR observations.  Overlaid on these images
  are the expected ultra high-magnification regions ($\mu > 100$) for
  a source at $z=9.2$ based on the gravitational lensing models we
  have for the three clusters (Z12; A. Zitrin et al.\ 2012, in prep;
  M. Carrasco et al.\ 2012, in prep).  Our lensing models for
  MACSJ1115.9+0129 and MACSJ1720.3+3536 are still preliminary and have
  not yet been finalized, constructed merely with the assumption that
  mass traces light, with typically only one lower-redshift system for
  normalization.  The position of our three candidates is indicated by
  the large magenta circles.  The dashed yellow circles indicate the
  position of possible counterimages as predicted by our preliminary
  lensing models.\label{fig:montage}}
\end{figure*}

\begin{figure*}
\epsscale{0.75}
\plotone{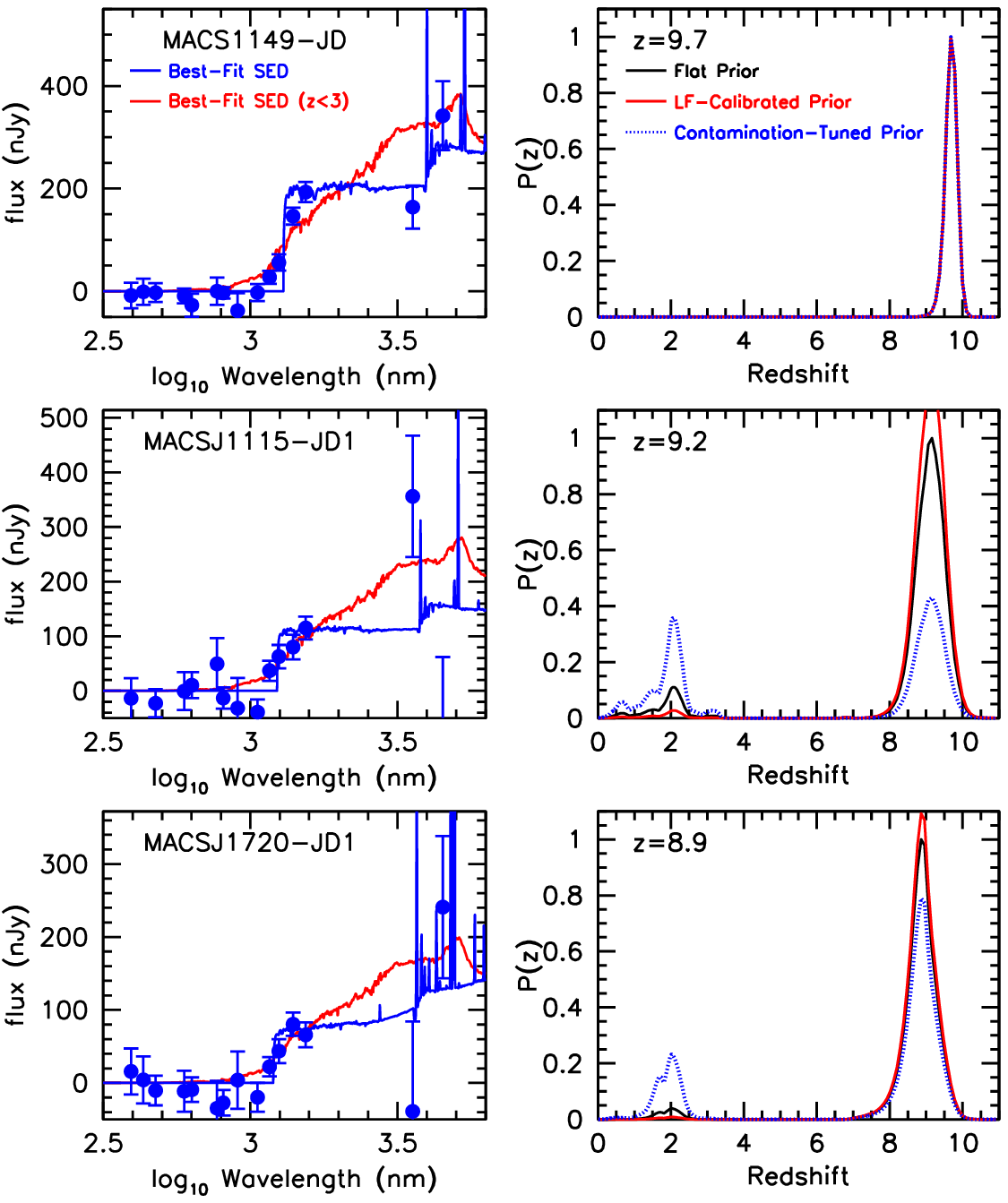}
\caption{(\textit{left}) Observed spectral energy distributions
  (\textit{solid blue circles}) for three $z\sim9$ galaxy candidates
  in our selection.  The blue line shows the SED template which best
  fits our observed photometry (using the EAZY photometric redshift
  code), while the red line shows the best-fit $z\sim0$-3 SED
  template.  The candidate in the uppermost row was previously
  presented in Z12.  (\textit{right}) The redshift likelihood
  distribution computed for our three $z\sim9$ candidates using the
  EAZY photometric redshift software (see \S3.4).  We consider three
  different priors in computing the redshift likelihood distributions:
  (1) a flat prior (\textit{black line}), (2) a prior calibrated to
  reproduce published LFs or LF trends (\textit{red line}: Giallongo
  et al.\ 2005; Bouwens et al.\ 2011b; R. Quadri et al.\ 2012, private
  communication), and (3) a prior tuned to reproduce the results from
  our photometric scattering simulations (\textit{dotted blue line}:
  \S3.5).  Appendix A provides a more detailed description of these
  priors.  Results from our third prior account for the fact that
  $\sim$1 faint source from our selection might be expected to
  resemble plausible $z\sim9$ galaxies, due to the effects of noise
  (see \S3.5).  However, even though we might expect a source to
  possibly scatter into our $z\sim9$ selection, we have no evidence
  that any particular source in our sample actually corresponds to
  such a low-redshift interloper.\label{fig:seds}}
\end{figure*}

\subsection{Resulting $z\sim9$ Sample}

We applied the selection criteria given in the previous section to the
HST WFC3/UVIS+ACS+WFC3/IR observations from all 19 clusters in the
current data set.  We identified three sources which satisfy these
selection criteria.  The sources are found behind three different
clusters MACSJ1149.6+2223, MACSJ1115.9+0129, and MACSJ1720.3+3536.
The brightest of our three candidates, i.e., MACS1149-JD, was
already presented in Z12.  These sources were also flagged as the most
interesting $z>8$ sources using an independent, purely photometric
redshift selection (Bradley et al.\ 2013).

In narrowing our selection down to our three highest quality
$z\sim9$-10 candidates, we found that our $\chi_{opt+Y}^2$ optical
non-detection and $JH_{140}-H_{160}<0.5$ criteria were particularly
important.  From the small sample of 29 sources that satisfied our
$(J_{125}+JH_{140})/2-H_{160}>0.7$ Lyman Break criterion and optical
non-detection criteria in individual bands, we found that our
$z\sim9$-10 reduced down to 9 sources if we applied our $\chi_{opt+Y}
^2$ criterion and finally down to our 3 candidates if we applied the
$JH_{140}-H_{160}<0.5$ color criterion.  Use of either of the two
$J$-band filters in constructing a Lyman-break sample resulted in a
similar selection of sources, modulo one or two sources.  For example,
application of $J_{110}-H_{160}>0.7$ color selection instead of the
$(J_{110}+J_{125})/2-H_{160}>0.7$ selection resulted in the same
$z\sim9$-10 candidates as are featured in our paper plus a source at
04:16:11.53, $-$24:04:53.2, which appears quite likely to be at
$z\sim8.4$ (i.e. just outside our redshift selection window).

We performed Spitzer/IRAC photometry on all three $z\sim9$-10
candidates using the software described in \S3.1.  None of the three
sources is nearby a bright foreground source and so all of our IRAC
flux measurements should be reliable.  Two of our three $z\sim9$-10
candidates (MACS1149-JD and MACS1720-JD1) are detected ($>2\sigma$) in
the moderately deep Spitzer/IRAC observations now available over
MACS1149 and MACS1720.

The coordinates and photometry of these candidates are provided in
Table~\ref{tab:properties}, while postage stamp images of the
candidates are shown in Figure~\ref{fig:stamps}.  In
Table~\ref{tab:properties}, we also present a mean spectral energy
distribution for galaxies at $z\sim9$, which we computed on the basis
of our HST+Spitzer photometry for the three $z\sim9$ candidates.  In
computing this mean SED, the fluxes of each source are rescaled such
that its average $JH_{140}$+$H_{160}$ flux matches the average
$JH_{140}+H_{160}$ flux for the sample (prior to rescaling).

As shown in Figure~\ref{fig:stamps}, MACS1149-JD is clearly resolved
(see the Supplementary Information to Z12).  MACS1149-JD also shows
distinct elongation along the shear axis (Figure 1 from Z12) predicted
from our gravitational lensing model for MACSJ1149.6+2223 (Z12).  The
other two plausible $z\sim9$ candidates in our selection are quite
small and show no clear evidence for gravitational shearing in the
expected directions.  However, since we would expect faint $z\geq9$
galaxies to be small and the predicted magnification to be only modest
(magnifications of $\sim$5-9$\times$ in total), it is not clear that
the structural properties of the sources teach us anything definitive.

In Figure~\ref{fig:montage}, we indicate the position of these
candidates within the field of view of our MACSJ1149.6+2223,
MACSJ1115.9+0129, and MACSJ1720.3+3536 observations (\textit{magenta
  circles}).  On Figure~\ref{fig:montage}, we have also overplotted
the approximate critical lines for these clusters based on the lens
models we have for these clusters (\textit{white contours}: Z12;
Zitrin et al.\ 2012, in preparation; Carrasco et al.\ 2012, in
preparation).  We caution that the lens models we have for
MACSJ1115.9+0129 and MACSJ1720.3+3536 are still somewhat preliminary
and are not totally finalized yet.  The models are constructed based
on the assumption that mass traces light, with typically only one
lower-redshift system for normalization.

We can use these magnification models to estimate the approximate
magnification factors for our candidate $z\sim9$ galaxies.  The
approximate magnification factors are 14.5$_{-1.0}^{+4.2}$,
9.3$_{-3.6}^{+5.8}$, and 5.0$_{-0.7}^{+4.7}$ and suggest intrinsic
delensed $H_{160,AB}$ magnitudes for the sources of 28.5, 28.6, and
28.6 mag, respectively, for MACS1149-JD, MACSJ1115-JD1, and
MACSJ1720-JD1.  The intrinsic magnitudes inferred for the first three
$z\sim9$ galaxy candidates in the CLASH sample are only slightly
brighter than was found for the Bouwens et al.\ (2011a) $z\sim10$
candidate, i.e., $H_{160,AB}\sim28.7$ mag, and seem consistent with
expectations.

The predicted positions of any possible counterimages to our $z\sim9$
candidates are also shown on Figure~\ref{fig:montage} (\textit{dashed
  yellow circles}).  The only case where the counterimages are
expected to be bright enough to detect is for MACSJ1720-JD1.
Unfortunately, we were unable to locate the counterimages to
MACSJ1720-JD1 at the predicted positions -- which could mean that our
lensing model may require further refinements, the counterimages are
blended with foreground sources, or that the redshift identification
is incorrect.  For MACSJ1115-JD1, the counterimage is expected to be
too faint to detect.

\subsection{Best-fit Photometric Redshifts}

The three candidate $z\sim9$ galaxies presented in the previous
section were selected using a two-color Lyman-Break selection, and
therefore their photometry is likely a reasonable fit to a model
star-forming galaxy SED at $z\sim9$.  However, since one can often fit
the same photometry with SED templates at different redshifts, it is
worthwhile for us to examine these candidates using standard
photometric redshift procedures to look for possible degeneracies.
Our use of photometric redshift procedures also allow us to naturally
fold in the IRAC flux information we have for our $z\sim9$ candidates.

To this end, we used the EAZY photometric redshift software (Brammer
et al.\ 2008) to estimate photometric redshifts for the sources based
on the observed photometry and to calculate the relative probability
that sources in sample are more likely star-forming galaxies at
$z\sim9$ or galaxies at lower redshift (i.e., $z<3$).  The photometric
redshift fitting is conducted using the EAZY\_v1.0 template set
supplemented by SED templates from the Galaxy Evolutionary Synthesis
Models (GALEV: Kotulla et al.\ 2009), which includes nebular continuum
and emission lines as described in Anders \& Fritze-v. Alvensleben
(2003).  The EAZY\_v1.0 template set consists of five SED templates
from PEGASE library (Fioc \& Rocca-Volmerange 1997) derived based on
the Blanton \& Roweis (2007) algorithm and one young, dusty template
(50 Myr, $A_V = 2.75$).

We consider three different priors in looking at the redshift
likelihood distribution of our three $z\sim9$ candidates: (1) a flat
prior, (2) a prior calibrated to published LFs or LF trends, and (3) a
prior tuned to reproduce the contamination rate estimated in the next
section (\S3.5).  Our second prior is based on the LF results of
Giallongo et al.\ (2005) and R. Quadri et al.\ (2012, private
communication) for red $z\sim1.3$-2 galaxies while at $z>7$ we utilize
the LF-fitting formula of Bouwens et al.\ (2011b).  The third prior
accounts for the effect of noise on the photometry of lower-redshift
galaxies in our search fields and the fact that in some rare events,
noise could cause $\sim$1-2 sources from our fields to seem like
highly probable $z\sim9$ galaxies (\S3.5).  Our third prior is
calibrated to reproduce the results from our photometric scattering
experiments.  For simplicity (and because of the similar luminosities
and magnitudes of all three of our $z\sim9$ candidates), these priors
are only a function of redshift; no luminosity dependence is
considered.  A more detailed description of these priors is provided
in Appendix A.

The results are shown in Figure~\ref{fig:seds}.  The left panels show
a comparison of the observed photometry with the best-fit $z\sim9$-10
galaxy (\textit{blue line}) and best-fit $z<3$ galaxy (\textit{red
  line}), while the right panels show the probability that a given
source in our sample has a particular redshift.  The best-fit
redshifts for MACS1149-JD, MACSJ1115-JD1, and MACSJ1720-JD1 using the
flat priors were 9.7, 9.2, and 8.9, respectively.  The 68\% confidence
intervals on the derived redshifts based on these same priors are
[9.57,9.78], [8.38, 9.57], and [8.38, 9.26], respectively.

No substantial changes in these results are seen using our other two
priors, except for the integrated probability within the $z\sim1$-2
peak.  For our second LF-calibrated prior (\textit{red line}), the
lower-redshift peak is actually smaller than in the case of the flat
prior.  This simply reflects the extreme rarity of faint red (old
and/or dusty) galaxies at $z\sim1.3$-2 as found in the Giallongo et
al.\ (2005) and R. Quadri et al.\ (2012, private communication) probes
(see also Stutz et al.\ 2008 and Figure 11 from Oesch et al.\ 2012a).
For our third prior (\textit{dotted blue lines}), the lower-redshift
peak is larger, particularly for MACSJ1115-JD1 and MACS1720-JD1.
Indeed, we might expect the lower-redshift peak to be higher than we
would estimate from the photometry (and a flat prior), due to the
impact that the selection process itself has on the observed SEDs of
sources that satisfy our selection criteria.  The selection process
itself picks out those particular noise realizations for individual
sources that are most consistent with those sources appearing
consistent with being $z\sim9$-10 galaxies (even if that is not
actually the case).

For the likelihood distributions given for the third prior, we should
emphasize that the likelihood distributions were tuned so as to
reproduce the expected contamination level for our $z\sim9$ selection
over the first 19 CLASH clusters (suggesting some possible
contamination of our selection by lower redshift interlopers) and that
we have no evidence that one particular source from our selection
(e.g., MACSJ1115-JD1 or MACSJ1720-JD1) is in fact a contaminant.

As many the filters in the CLASH program have overlapping wavelength
coverage, we can further test the robustness of our best-fit
photometric redshifts by making using of either the $J_{110}$ or
$J_{125}$ flux measurements and making use of either the $JH_{140}$ or
$H_{160}$ flux measurement.  The best-fit redshifts we find for
MACS1149-JD, MACS1115-JD1, and MACS1720-JD1 range from 9.6 to 9.8, 9.1
to 9.1, and 9.0 to 9.4, respectively.  If we exclude the $Y_{105}$
flux measurement in deriving the best-fit photometric redshift, we
find similar photometric redshifts for MACS1149-JD and MACS1720-JD1,
but find a best-fit photometric redshift of 1.2 for MACS1115-JD1.  

We also derived redshift likelihood confidence intervals using the Le
PHARE photometric redshift package (Arnouts et al.\ 1999; Ilbert et
al.\ 2006, 2009) for our three candidates.  The SED templates we used
with Le PHARE were the same ones as optimized for the COSMOS survey
(Scoville et al.\ 2007) and included three elliptical and six spiral
SEDs as generated by Polletta et al.\ (2007) using the GRASIL code
(Silva et al.\ 1998) as well as 12 starburst galaxies ranging in age
from 30 Myr to 3 Gyr using the Bruzual \& Charlot (2003) GALAXEV
library.  We supplemented these with four additional elliptical
templates for a total of seven elliptical templates.  Dust extinction
was added in ten steps up to $E(B-V)=0.6$.\footnote{Of course,
  allowing for an even larger range of reddenings would be useful for
  more fully considering the possibility these candidates might
  correspond to ULIRGs.  However, the moderately blue colors of our
  three candidates likely rules out this possibility.} With these
templates, we used Le PHARE to derive the following 68\% confidence
intervals for the candidates: [9.49,9.85] for MACS1149-JD, [8.77,9.57]
for MACSJ1115-JD1, and [8.65,9.31] for MACSJ1720-JD1.  The best-fit
redshifts for these three candidates were 9.68, 9.17, and 8.93,
respectively.  The above results are for a flat prior and are quite
similar to what we derived using EAZY.  Use of the two other priors
resulted in similar changes to the redshift likelihood distributions
as shown in Figure~\ref{fig:seds}.

Finally, we also estimated the photometric redshifts of our three
candidates with BPZ (Bayesian Photometric Redshift Code:
Ben{\'{\i}}tez 2000; Coe et al.\ 2006).  Similar to the analyses in
C13 and Z12, we modelled the photometry using SEDs from PEGASE (Fioc
\& Rocca-Volmerange 1997) adjusted and recalibrated to match the
observed photometry of galaxies with known spectroscopic redshifts in
the FIREWORKS catalog (Wuyts et al.\ 2008).  This FIREWORKS catalog
includes photometry to 24.3 AB mag ($5\sigma$) in $K_s$, for galaxies
with $z\sim3.7$.  The best-fit photometric redshifts we derive with
BPZ are 9.7, 9.2, and 8.9 for MACS1149-JD, MACSJ1115-JD1, and
MACSJ1720-JD1, respectively.  For MACS1149-JD, the redshift likelihood
distribution is predominantly uni-modal though in the other two cases
the distribution is more bimodal, with modest peaks at lower redshift.
29\% and 5\% of the total probability for MACS1115-JD1 and
MACS1720-JD1, respectively, is at $z\sim1.0$-2.0.  Focusing on the
dominant $z\sim9$ peaks (excluding all $z<5$ solutions), the 68\%
confidence intervals on the redshifts for our three candidates are
[9.56, 9.87], [8.45, 9.55], and [8.30,9.26], respectively.  These
results are for a flat prior and are somewhat similar to what we
derived using the other two photometric redshift codes, although the
low-redshift peaks are slightly more significant with BPZ.  We opted
not to make use of the BPZ prior in computing the redshift
distribution for our sources, due to the relative weight it assigns to
faint red galaxies at $z\sim1.3$-2 and blue galaxies at $z\sim9$
(which differs by more than a factor of 30 from what we compute based
on published LFs or LF trends: see Appendix A).  We find that a flat
prior comes much closer to accurately representing the relative
surface densities of these two populations.

\begin{figure}
\epsscale{1.15}
\plotone{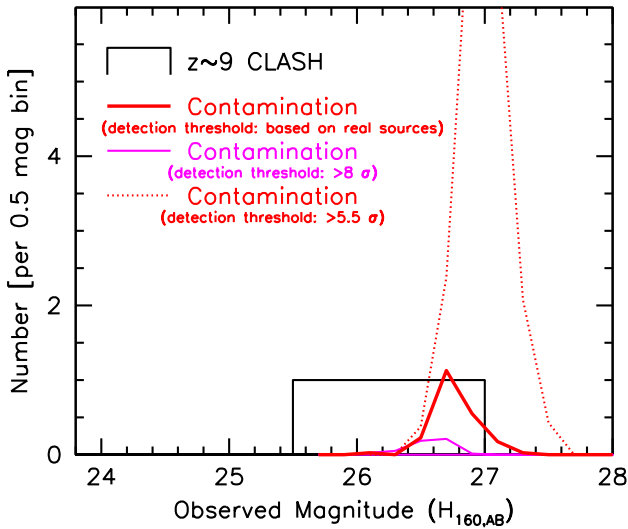}
\caption{The number of $z\sim9$ galaxy candidates we find in our CLASH
  cluster search as a function of the $H_{160,AB}$-band magnitude.
  Also plotted (\textit{red line}) is the number of contaminants we
  would expect to select in our search fields for sources, due to the
  effects of noise on the photometry of other lower redshift sources
  in search fields (see \S3.5 for details).  The total number that we
  estimate for our search fields is 0.7 (versus the 3 $z\sim9$
  candidates in our selection).  In modeling possible contamination of
  our selection, we only allow for three contaminants at maximum and
  the \textit{n}th contaminant must have a higher signal to noise than
  the \textit{n}th lowest signal-to-noise source.  For context, we
  also show the contamination expected for a $>$8$\sigma$ selection
  and for a $>$5.5$\sigma$ selection.  Clearly, contamination from
  lower redshift sources (due to photometric scatter) is only
  especially significant for sources with $H_{160,AB}$-band magnitudes
  faintward of 26.5 AB mag.  For sources detected at just $5.5\sigma$
  in the $JH_{140}+H_{160}$ bands (with magnitudes $\sim$27 AB mag),
  contamination from lower redshift becomes very
  important.\label{fig:surfdens}}
\end{figure}

\subsection{Possible Contamination}

While the sources in our current selection are consistent with being
$z\sim9$ galaxies, these sources are faint enough that they could
easily have a very different nature.  Important sources of
contamination for $z>8$ selections include low-mass stars, supernovae,
emission line galaxies (van der Wel et al.\ 2011), and the photometric
scatter of various low-redshift galaxies.  Readers are referred to
Bouwens et al.\ (2011a), Z12, and C13 for rather extended discussions
of these issues.

In general, the most important source of contamination for
high-redshift samples results from faint Balmer-break galaxies
entering these samples (and hence satisfying their selection criteria)
due to the effects of noise (see discussion in Wilkins et al.\ 2011,
Bouwens et al.\ 2011, Bouwens et al.\ 2014b).  Noise can cause such
galaxies (with faint optical flux and not especially red) to look
bluer and disappear entirely at optical wavelengths.

Here we test for contamination from faint lower-redshift sources
scattering into our high-redshift selection through the effects of
noise, by using all intermediate magnitude sources in the CLASH
cluster fields that are detected at $>2\sigma$ in the $I_{814}$ band
and $Y_{105}$ bands (and therefore likely at redshifts $z<6$) to
implicitly define the color distribution for potential interlopers to
our high-redshift samples.  Then, we take all the faint sources in all
the CLASH cluster fields (with their $H_{160,AB}$ magnitudes and
errors), randomly match them up with a source from the sample which
defines our color distribution, give this faint source the same colors
as the intermediate-magnitude source, add noise to the photometry of
the sources in its bluer bands (assuming a normal distribution), and
then see if this source satisfies our $z\sim9$ selection criteria.
Our procedure here is essentially identical to what we performed in
many previous analyses (e.g., Bouwens et al.\ 2011a; Bouwens et
al.\ 2011b).  In modeling possible contamination of our selection, we
only allow for three contaminants at maximum and the \textit{n}th
contaminant per CLASH data set must have a higher signal to noise than
the \textit{n}th lowest signal-to-noise source.

\begin{figure}
\epsscale{1.15}
\plotone{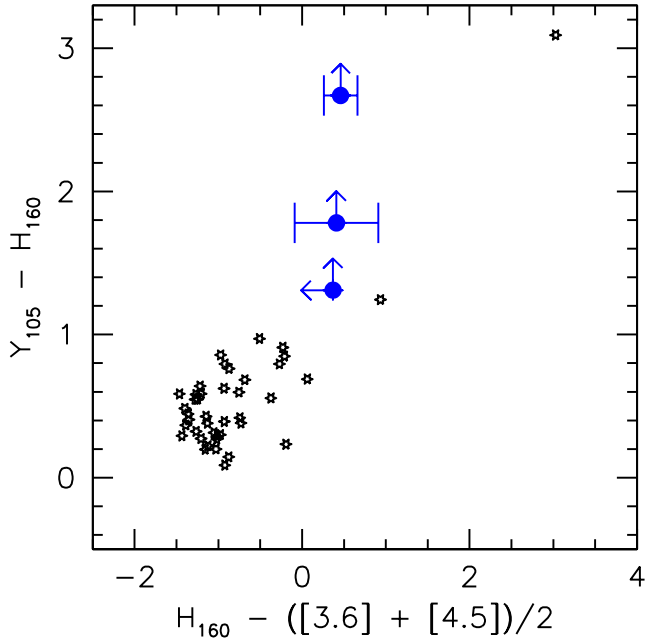}
\caption{A comparison of the $Y_{105}-H_{160}$
  vs. $H_{160}-([3.6]+[4.5])/2$ colors of the three $z\sim9$
  candidates in our sample (\textit{solid blue circles},
  \textit{$1\sigma$ error bars}, and \textit{arrows indicating
    $1\sigma$ limits}) with the observed colors of various stars.  The
  black starlike symbols are the colors derived from the substantial
  library of stellar spectra observed with IRTF (Cushing et al.\ 2005;
  Rayner et al.\ 2009), with sources ranging from very low-mass stars
  to higher mass Mira-type variable stars (\textit{the black starlike
    symbol in the upper right of this figure}).  Two of our $z\sim9$
  candidate galaxies have $Y_{105}-H_{160}$ colors which are clearly
  too red to match those colors observed by the broad set of stars
  encompassed by this library.\label{fig:star}}
\end{figure}

Applying this procedure to all the sources in the CLASH fields
100$\times$, we find that only 0.7 lower-redshift ($z\lesssim6$)
sources enter our $z\sim9$ selection by chance (per Monte-Carlo
simulation for the entire CLASH program).  The magnitude distribution
of these contaminants is shown in Figure~\ref{fig:surfdens}.  The
small number of contaminants we find from these simulations
demonstrate that the overall level of contamination for the present
probe is likely only modest ($\sim$23\%), becoming important faintward
of 26.5 mag.

We also considered the implications for contamination if we had
restricted our selection to sources with a $JH_{140}+H_{160}$
detection significance of $>$8$\sigma$ and $>$5.5$\sigma$ (weaker than
our $z\sim9$ selection criteria).  The results are shown in
Figure~\ref{fig:surfdens} with the magenta and dotted red lines,
respectively.  Only $\sim$0.2 contaminants are expected for a
$8\sigma$ detection threshold while for a $\sim$5.5$\sigma$ threshold
the expected number of contaminants is $\sim$7.5 sources -- and hence
might be a significant concern if we had considered a lower detection
threshold for our selection.

This being said, it is worth noting that our estimate of the total
contamination here may be a little high (perhaps by a factor of
$\sim$2-3), due to our use of an intermediate magnitude
($\sim$24.0-25.5 mag) population of galaxies to model the colors of
somewhat fainter galaxies (i.e., 26.0-27.0 mag).  Since the
intermediate magnitude population are somewhat redder in general than
$\sim$26.0-27.0 mag population (e.g., see Figure 11 of Oesch et
al.\ 2012a), they are more likely to scatter into $z\sim9$ selections
via noise than is the actual situation for $\sim$26.0-27.0 mag
galaxies.

In any case, the results of these simulations strongly suggest that
the two most significantly detected sources in our sample, i.e.,
MACSJ1115-JD1 and especially MACS1149-JD, are unlikely to correspond
to such contaminants.  For sources with lower S/N than this, we must
remain concerned about contamination -- even though we cannot
establish the exact rate.  The issue will contribute to the overall
errors in our SFR density estimates at $z\sim9$ (\S4).

Of course, faint moderately blue low-redshift galaxies are not the
only galaxies that can contaminate high-redshift samples.  Dust
reddened galaxies can also occasionally contaminate high-redshift
selections (Bowler et al.\ 2012; Laporte et al.\ 2011), as well as
lower-redshift galaxies with somewhat unusual SEDs (Hayes et
al.\ 2012; Boone et al.\ 2011).  While it is difficult to be sure that
sources in our sample do not correspond to such galaxies at lower
redshift, all three sources in our samples generally have bluer colors
than those lower-redshift contaminants, and so we suspect such sources
do not pose a problem for our selection.  The moderately red color of
MACS1720-JD1 in the $3.6\mu$m and $4.5\mu$m bands is similar to the
colors seen in other $z\sim8$ candidates (e.g., Ono et al.\ 2012;
Labb{\'e} et al.\ 2013; Finkelstein et al.\ 2013; Laporte et
al.\ 2014) likely showing strong \textsc{[O\,iii]} emission.

Another possible source of contamination is from extreme emission-line
galaxies (EELGs) with strong \textsc{[O\,iii]}+H$\beta$ emission, such
as recently discovered by van der Wel et al.\ (2011) and Atek et
al.\ (2011) in the CANDELS (Grogin et al.\ 2011; Koekemoer et
al.\ 2012) or WISP (Atek et al.\ 2010) programs.  Perhaps the most
well-known high-redshift candidate thought to be such an EELG is the
Bouwens et al.\ (2011) $z\sim2$/$z\sim12$ candidate UDFj-39546284
(Ellis et al.\ 2013; Brammer et al.\ 2013; Bouwens et al.\ 2013; Capak
et al.\ 2013).  However, it seems unlikely that any of our sources
correspond to such candidates given that their detection in multiple
non-overlapping bands and blue $UV$-continuum slope of most EELGs (van
der Wel et al.\ 2011).

Finally, there is the possibilities that candidates from our selection
could correspond to stars or supernovae (SNe).  Both possibilities
would require that sources in our selection are unresolved.  Comparing
the coadded $JH_{140}+H_{160}$ profile of our candidates with the
WFC3/IR PSF, it is clear that 2 out of our 3 candidates are resolved
(see also discussion in Z12 which demonstrate clearly that MACS1149-JD
is resolved).  Only MACSJ1115-JD1 does not show any spatial extension.
In any case, as Figure~\ref{fig:star} demonstrates, the colors of the
candidates do not clearly support a stellar origin.  The redder
Mira-variable stars would appear to give the best match, but their
intrinsic luminosities are such that we would need to observe them
well outside our own Milky Way galaxy (Whitelock et al.\ 1995;
Dickinson et al.\ 2000).  We can also safely exclude the possibility
of a SNe, given that deep optical observations of our cluster fields
were obtained over the same two month time window as our deep near-IR
observations (Postman et al.\ 2012; see also Z12 and C13).

One final possibility is that some candidates may correspond to more
local solar system or Oort cloud objects.  To be consistent with the
constraints we can set on the proper motion of our candidates based on
the $\sim$2 month observational baseline we have (see Figure 6 of C13
for an illustration of the constraints we can set), such a source
would need to be at 50,000 AU.  However, at such distances, Oort cloud
objects would be extremely faint (e.g., faintward of 40 mag), even if
as these sources were as large as the moon (see also discussion in Z12
and C13).

\section{A New Differential Determination of the $UV$ Luminosity Function
at $z\sim9$}

The present $z\sim9$ sample is the largest such sample available to
date and should allow us to substantially improve our constraints on
the $z\sim9$ luminosity function.  However, before providing a
detailed discussion of the specific constraints we are able to set, we
must first include a few words on the procedure we adopt.

\subsection{$UV$ LF Evolution from Lensing Cluster Searches: Rationale for Using a Differential Approach}

Normally, we would derive the luminosity function for $z\sim9$
galaxies using the same approach that has been followed in the field,
i.e., (1) distribute the sources in one's samples into different
magnitude intervals, (2) count the total number of sources in a
magnitude interval (after correcting for contamination) and (3) divide
these numbers by the effective volume where such sources could be
found.  Such a procedure has been followed in a number of previous
works (e.g., Santos et al.\ 2004; Richard et al.\ 2008).

However, even the simple process of placing sources into different
intrinsic magnitude bins can be quite uncertain due to its dependence
on a particular magnification model.  Calculations of the selection
volumes are just as equally model dependent.  While in many cases
these model dependencies may not result in large overall uncertainties
in one's results, the uncertainties clearly do become large
($\sim$0.3-0.4 mag or larger) near the critical curves of the lensing
models where the magnification factors become nominally infinite
(e.g., see Figure 2 of Maizy et al.\ 2010).  These issues can
potentially have a huge effect on luminosity functions derived in the
context of lensing clusters.\footnote{Of course, for realistic LFs,
  these uncertainties may not be especially problematic.  Indeed, for
  LFs with an effective faint-end slope close to $-2$, uncertainties
  in the magnification factor trade off almost perfectly with
  uncertainties in the search volume so as to have no large effect on
  the inferred LFs.  Because of this fact, one potentially very
  effective approach for minimizing the impact these uncertainties on
  the derived LFs is by marginalizing over the magnification factor in
  performing the comparisons with the observed numbers (C13).  The
  excellent agreement between the present estimate of the SFR density
  at $z\sim9$ and that obtained by C13 based on the Z12 search results
  would seem to support this conclusion.}

\begin{figure}
\epsscale{1.15}
\plotone{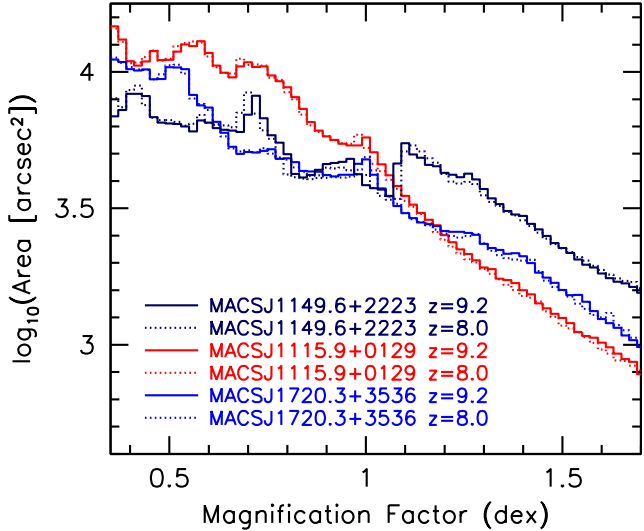}
\caption{Search area (per unit dex) behind select galaxy clusters
  subject to varying levels of magnification by gravitational lensing.
  Results are shown for sources at $z=8$ and $z=9$ based on the lens
  models for MACSJ1149.6+2223, MACSJ1115.3+0129, and MACSJ1720.3+3536
  (Z12; Zitrin et al.\ 2012, in prep; Carrasco et al.\ 2012, in prep).
  It is obvious from these results that the total search volume behind
  a cluster (given the area magnified to various levels) can show a
  huge variation from one cluster to another.  However, if one
  utilises the same cluster to search for sources at similar but
  slightly different redshifts (\textit{compare the dotted and solid
    lines representing $z\sim8$ and $z\sim9$ selections}), almost
  exactly the same selection area is available for selecting sources
  at a given magnification factor (although we remark that the
  selection area is slightly larger ($\sim$1-3\%) at $z\sim9$ than at
  $z\sim8$).  As a result, we would expect the relative selection
  volumes for a $z\sim9$ search behind lensing clusters and a $z\sim8$
  search behind lensing clusters to be very well defined, \textit{if
    the same clusters are utilized for the two
    searches}.\label{fig:relvol}}
\end{figure}

\begin{figure*}
\epsscale{1.15}
\plotone{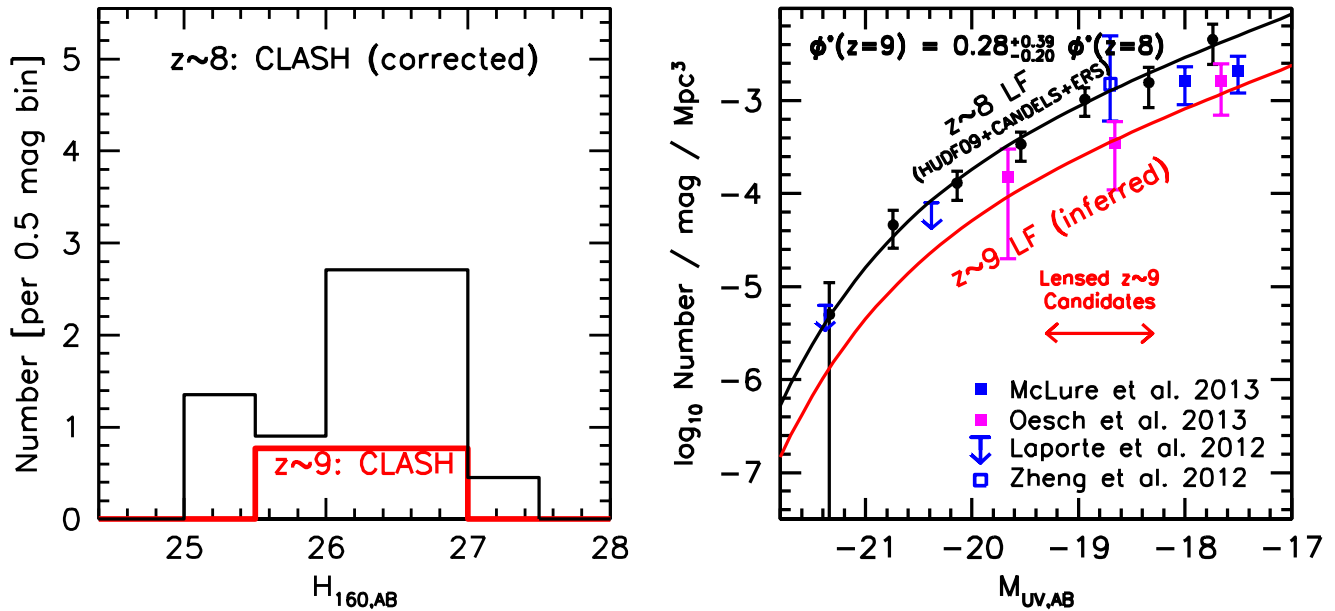}
\caption{Illustration of our differential approach to deriving the
  $UV$ LF at $z\sim9$.  (\textit{left}) The contamination-corrected
  number of $z\sim9$ galaxy candidates we find within CLASH
  (\textit{red histogram}) vs. the number of $z\sim8$ galaxy
  candidates (\textit{black histogram}) behind the same CLASH
  clusters, corrected to have the same selection volume as at
  $z\sim9$.  For simplicity, the contamination rate correction is
  applied in a magnitude-independent manner (although the
  contamination rate will clearly be higher near the faint ends of our
  two samples).  The number of $z\sim9$ galaxy candidates in CLASH,
  after contamination correction, is just $0.28_{-0.20}^{+0.39}\times$
  that at $z\sim8$.  A simple comparison of these surface densities
  should give us a fairly model independent measure of the relative
  normalization of the $UV$ LF at $z\sim8$ and the $UV$ LF at $z\sim9$
  -- assuming that the shape of the LF (i.e., $M^*$ and $\alpha$) does
  not change very dramatically from $z\sim9$ to $z\sim8$.
  (\textit{right}) The observed $UV$ LF at $z\sim8$ as derived by
  Oesch et al.\ (2012b: \textit{black points, error bars, and line})
  based on the HUDF09+CANDELS+ERS data set and our newly inferred $UV$
  LF at $z\sim9$ (\textit{red line}) based on our differential
  comparison of our $z\sim8$ and $z\sim9$ selections.  We infer that
  the $UV$ LF at $z\sim9$ has an effective $\phi^*$ that is just
  $0.28_{-0.20}^{+0.39}\times$ that at $z\sim8$.  The red horizontal
  arrow towards the bottom of this panel indicates the approximate
  luminosities inferred for our 3 $z\sim9$ candidates (after
  correction for lensing magnification: see \S3.3).  For context, we
  also show recent constraints on the volume density of $z\sim9$
  galaxies from Zheng et al.\ (2012), Laporte et al.\ (2012), McLure
  et al.\ (2013), and Oesch et al.\ (2013).\label{fig:newlf}}
\end{figure*}

\begin{deluxetable}{ccccc}
\tablewidth{0pt} \tabletypesize{\footnotesize}
\tablecaption{Estimated Schechter Parameters for the $UV$ LF at $z\sim9$ and a comparison with $UV$ LF determinations at other redshifts $z\sim4$-10 (see \S4.4).\label{tab:lfparm}}
\tablehead{ \colhead{Dropout} & & & \colhead{$\phi^*$ ($10^{-3}$}
  \\ \colhead{Sample} & \colhead{Redshift} & \colhead{$M_{UV}
    ^{*}$\tablenotemark{a}} & \colhead{ Mpc$^{-3}$)} &
  \colhead{$\alpha$}}
\startdata 
$J_{110}+J_{125}$ & 9.2 & $-20.04$ (fixed) & $0.14_{-0.11}^{+0.20} $ & $-2.06$ (fixed) \\ 
\multicolumn{5}{c}{-----------------------------------------------------------------} \\ 
$B$ & 3.8 & $-21.07\pm0.08$ & $1.41_{-0.20}^{+0.23}$ & $-1.64\pm0.04$\\ 
$V$ & 4.9 & $-21.19\pm0.11$ & $0.64_{-0.12}^{+0.14}$ & $-1.78\pm0.05$\\
$i$ & 5.9 & $-21.16\pm0.20$ & $0.33_{-0.10}^{+0.15}$ & $-1.91\pm0.09$\\
$z$ & 6.8 & $-21.04\pm0.26$ & $0.22_{-0.09}^{+0.14}$ & $-2.06\pm0.12$\\
$Y$ & 8.0 & $-20.04_{-0.48}^{+0.44}$ & $0.50_{-0.33}^{+0.70}$ &
$-2.06_{-0.28}^{+0.35}$\\
\enddata

\tablenotetext{a}{Values of $M_{UV}^{*}$ are at $1600\,\AA$ for the
  Bouwens et al.\ (2014b) $z\sim4$, $z\sim5$, $z\sim6$, and $z\sim7$
  LFs and at $\sim1750\,\AA$ for the Oesch et al.\ (2012b) constraints
  on the $z\sim8$ LF.  While the $z\sim8$ LF results from Bouwens et
  al.\ (2014) likely represent an improvement on those from Oesch et
  al.\ (2012b), we quote the Oesch et al.\ (2012b) results here since
  those represent our baseline for extending the LF results to
  $z\sim9$ (to maintain consistency with our earlier submission and
  because of excellent agreement between the Oesch et al.\ 2012b LF
  results and subsequent work at $z\sim8$).}
\end{deluxetable}

Ideally, we would like to determine the $UV$ LF at $z\sim9$ in a way
that avoids these uncertainties.  One possible way for us to do this
is (1) to leverage existing well-determined LFs that already exist at
$z\sim7$-8 from blank field studies (e.g., Bouwens et al.\ 2011b;
Oesch et al.\ 2012b; Bradley et al.\ 2012) not subject to potentially
large selection volume uncertainties and (2) then to use our searches
for $z\sim7$-10 galaxies behind lensing clusters to derive the
differential evolution in the LF from $z\sim9$ to $z\sim7$-8.  This
provides us with a somewhat indirect approach to deriving the LF at
$z\sim9$ and takes advantage of the very similar effect gravitational
lensing from low-redshift clusters has on light from the high-redshift
universe, regardless of the exact redshift of the source.
Fundamentally, this is due to the fact that the $D_{LS}/D_S$ factor is
very insensitive to redshift when the lensed source is at $z>5$ (i.e.,
very distant) and the lensing cluster is relatively close (i.e.,
$z\sim0.1$-0.5).  For example, for a $z\sim0.4$ lensing cluster, the
computed $D_{LS}/D_{S}$ factor for $z\sim9$ background sources is only
$\sim$1\% higher than the $D_{LS}/D_{S}$ factor for $z\sim8$ sources.
$D_{LS}$ and $D_S$ are the angular-diameter distances from the cluster
lens to source and from observer to source, respectively (e.g.,
Narayan \& Bartelmann 1996).

As a result, for sources seen behind a given lensing cluster, the
$z\sim8$ universe is magnified in almost exactly the same way as the
$z\sim9$ universe.  This can be illustated using the lensing models we
have available for three of the CLASH clusters
(Figure~\ref{fig:relvol}).  The total area available behind a given
cluster to magnify the background light by more than a factor of 3 is
almost exactly the same for the $z\sim8$ universe as for the $z\sim9$
universe.  Note that this is true, even if the precise position of the
critical curves at $z\sim9$ lies in a slightly different position from
the critical curves at e.g. $z\sim8$.

Because of the very similar way a given set of clusters magnifies
galaxies at $z\sim9$ and at other similar redshifts (e.g., $z\sim8$),
one might expect it to have the same effect on the total surface
densities of these galaxies one finds on the sky.  Therefore, if one
starts with the same luminosity function of galaxies at both $z\sim8$
and $z\sim9$, one would expect to find roughly the same surface
density of these galaxies on the sky, modulo two slight differences.
The $z\sim9$ galaxy distribution would be shifted to slightly fainter
magnitudes (e.g., by $\sim$0.3 mag versus $z\sim8$) to reflect their
slightly larger luminosity distances and would be present at slightly
lower surface densities (by $\sim$10\% versus $z\sim8$) reflecting the
smaller cosmic volume available at $z\sim9$.

Even multiple imaging of the same high-redshift sources would not
appreciably affect the ratio of sources seen at different redshifts,
since one would expect galaxies at $z\sim9$ and similar redshifts to
give rise to lensed multiplets to approximately the same degree, and
therefore the ratio of surface densities should be preserved.
However, since multiple images of a single background source are not
independent events, not accounting for this effect could have a slight
effect on the uncertainties we estimate for the relative surface
densities of galaxies at different redshifts.

Given this situation, it seems quite clear we should be able to use
the relative surface densities of galaxies in different redshift
samples to make reasonably reliable inferences about the relative
volume densities of the galaxy population at different epochs (after
making small adjustments to the numbers to account for the factors
discussed above).

\subsection{$z\sim8$ Comparison Sample}

Redshift $z\sim8$ selections serve as the perfect comparison sample
for our $z\sim9$ studies.  Not only is the $z\sim8$ universe close
enough to $z\sim9$ to make differences in the lensing effects quite
small overall, but the $\sim$70-80 $z\sim8$ galaxies available in
current WFC3/IR surveys allow the LF there to be robustly established
from field studies (e.g., Bouwens et al.\ 2011b; Lorenzoni et
al.\ 2011; Oesch et al.\ 2012b; Bradley et al.\ 2012).  This allows us
to put together the new information we have on the differential
evolution of the LF from $z\sim9$ to $z\sim8$ with previous $z\sim8$
LF determinations to estimate the approximate $UV$ LF at $z\sim9$.

Finally, given the observed rate of evolution in $M^*$ and $\alpha$
(e.g., using the fitting formula for Schechter [1976] parameterization
given in Bouwens et al.\ 2014b), we would expect the shape of the LF at
$z\sim8$ to be similar to the shape of the LF at $z\sim9$, i.e.,
$\Delta (M^* (z=8) - M^* (z=9)) \lesssim 0.2$ and $\Delta \alpha (z=8)
- \alpha (z=9) \lesssim 0.12$, so we can model any evolution in the LF
very simply assuming a change in the normalization $\phi^*$ (though
modeling the evolution in terms of the characteristic luminosity $M^*$
is only slightly more involved).

For our $z\sim8$ comparison sample, we use the same selection criteria
as previously utilized in Bouwens et al.\ (2011b) and Oesch et
al.\ (2012b), i.e.,
\begin{displaymath}
(Y_{105} - J_{125} > 0.45)\wedge(J_{125}-H_{160} < 0.5)
\end{displaymath}
As in these two previous works (and as performed for our $z\sim9$-10
selection), we also require sources to be undetected in the $I_{814}$
band and blueward both in individual bands at $<$2$\sigma$ and using
the $\chi_{opt} ^2$ statistic discussed earlier (\S3.2).  We also
demand that sources be detected at $>6\sigma$ in a combined $JH_{140}$
and $H_{160}$ image (0.35$''$-diameter aperture), as performed for our
primary $z\sim9$-10 selection.  Since these color criteria and
selection criteria are very similar to that used by Bouwens et
al.\ (2011b) and Oesch et al.\ (2012b) in identifying $z\sim8$
galaxies, the redshift distribution for the present $z\sim8$ selection
should be approximately the same as shown in Figure~\ref{fig:zdist}
(\textit{red line}).

Applying this selection criteria to the 19-cluster CLASH dataset, we
find a total of 19 sources which satisfy our $z\sim8$ criteria.  After
excluding one candidate from the sample (19:31:48.7, $-$26:34:03.0)
that is completely unresolved in the HST data\footnote{Median
  SExtractor stellarity parameter for this candidate is 0.94 in the
  $Y_{105}J_{110}J_{125}JH_{140}H_{160}$ data [where 1 and 0
    corresponds to a point and extended source, respectively].}  and
has colors very similar to that of low-mass stars, we are left with a
total sample of 18 $z\sim8$ candidates.  These sources have
$H_{160,AB}$ magnitudes ranging from 25.0 to 27.3 mag.  Coordinates of
these candidates and their $H_{160}$-band magnitudes are provided in
Table~\ref{tab:samplez8} from Appendix C.  We allow for a potential
contamination of $\sim$1.5 source in our $z\sim8$ sample, consistent
with the contamination level found by Bouwens et al.\ (2011b) for
their $z\sim8$ sample and also allowing for some possible
contamination by low-mass stars in our search fields.

The current $z\sim8$ selection includes more $z\sim8$ candidates per
cluster as the $z\sim8$ selection from Bradley et al.\ (2014) using
photometric redshifts over the same magnitude range.  This is due to
the present color criteria identifying galaxies at $z\gtrsim7.2$,
while the Bradley et al.\ (2014) photometric redshift criteria only
identify galaxies at $z\gtrsim 7.5$.  Our choice of selection criteria
should have little impact on our LF results, as the selection volumes
we compute for our $z\sim8$ sample (\S4.3) will largely offset any
changes in sample size.

\begin{figure*}
\epsscale{1.15}
\plotone{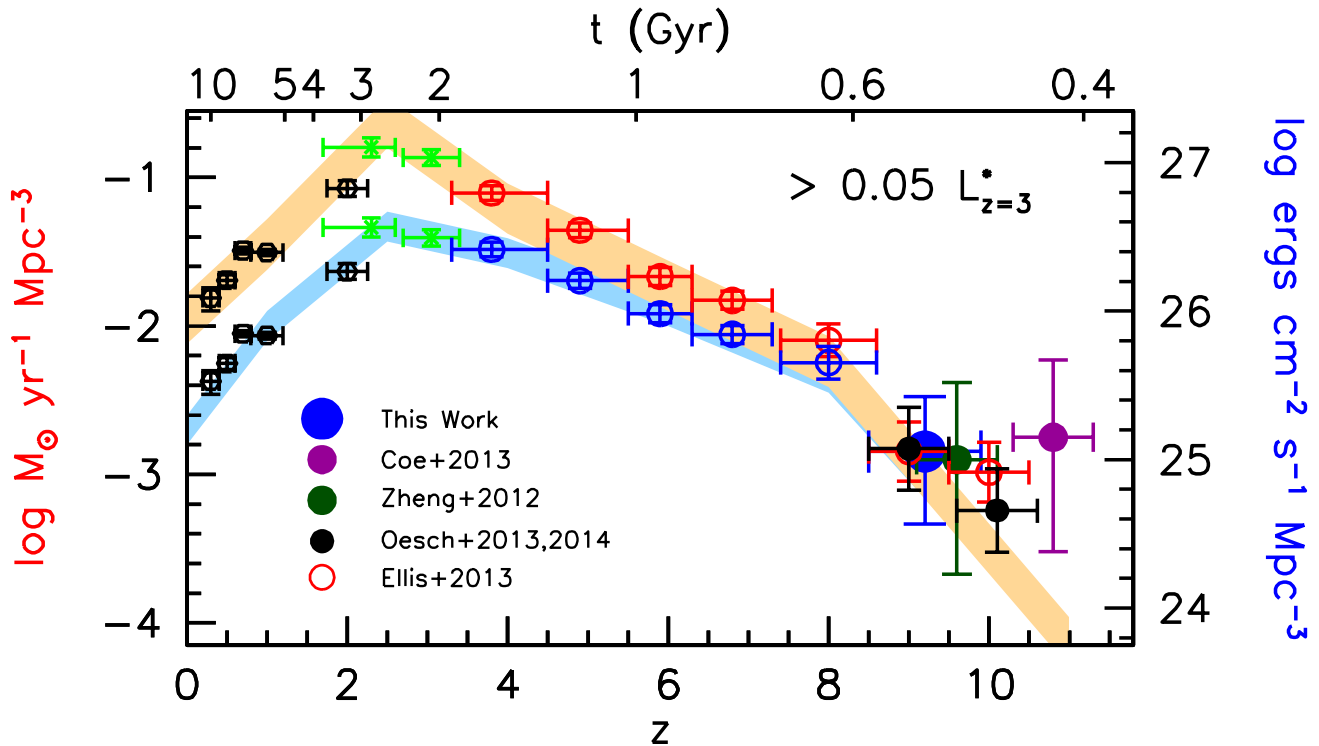}
\caption{The $UV$ luminosity density (\textit{right axis}) and star
  formation rate density (\textit{left axis}) versus redshift.  The
  $UV$ luminosity and SFR density shown at $z\sim9$ (\textit{large
    blue solid circle}) are from the present work and inferred based
  on the relative number of $z\sim8$ and $z\sim9$ galaxies found
  within the CLASH cluster program (see \S4.5).  These luminosity
  densities and SFR densities are only considered down to a limiting
  luminosity of $-$17.7 AB mag -- which is the approximate limit of
  both the HUDF09 probe (Bouwens et al.\ 2011b) and the present search
  assuming a maximum typical magnification factor of $\sim$9 and
  limiting magnitude of $\sim$27.0 mag.  The $UV$ luminosity is
  converted into a star formation rate using the canonical $UV$-to-SFR
  conversion factors (Madau et al.\ 1998; Kennicutt 1998).  The upper
  set of points at every given redshift and orange contour show the
  dust-corrected SFR densities, while the lower set of points and blue
  contours show the inferred SFR densities before dust correction.
  Dust corrections at $z>3$ are estimated based on the observed
  $UV$-continuum slope distribution and are taken from Bouwens et
  al.\ (2012b).  At $z\leq3$, the dust corrections are from
  Schiminovich et al.\ (2005) and Reddy \& Steidel (2009).  $UV$
  luminosity density and SFR density determinations from the
  literature are from Schiminovich et al.\ (2005) at $z<2$
  (\textit{black hexagons}), Reddy \& Steidel (2009) at $z\sim2$-3
  (\textit{green crosses}, Bouwens et al.\ (2007) at $z\sim4$-6
  (\textit{open red and blue circles}), Bouwens et al.\ (2011b) at
  $z\sim7$ (\textit{open red and blue circles}), Oesch et al.\ (2012b)
  at $z\sim8$ (\textit{open red and blue circles}), Ellis et
  al.\ (2013) at $z\sim9$-10 (\textit{open red circles}), Oesch et
  al.\ (2013) at $z\sim9$ (\textit{solid black circles}), and Oesch et
  al.\ (2014) at $z\sim10$ (\textit{solid black circles}).  Estimates
  of the SFR density at $z\sim9.6$ and $z\sim10.8$ as derived in C13
  based on the $z\sim9.6$ Z12 and $z\sim10.8$ C13 candidates are also
  shown (\textit{dark green and magenta solid circles, respectively}).
  Conversion to a Chabrier (2003) IMF would result in a factor of
  $\sim$1.8 (0.25 dex) decrease in the SFR density estimates given
  here.  The present $z\sim9$ determination is in good agreement with
  the trend in the SFR density and $UV$ luminosity, as defined by the
  Oesch et al.\ (2012a) and Z12 estimates.\label{fig:sfrdens}}
\end{figure*}

\subsection{Relative selection volumes at $z\sim8$ and $z\sim9$: Expected sample sizes assuming no evolution}

In order to utilize the relative surface density of $z\sim8$ and
$z\sim9$ galaxy candidates we observe to make inferences about the
evolution of the luminosity function, we must have an estimate for how
many galaxies we would expect in the two samples if the $UV$ LF did
not evolve at all between the two epochs.  Then, based on the relative
number of sources expected in the two samples assuming no evolution,
we can determine the approximate evolution in the LF from $z\sim9$ to
$z\sim8$.  With this step, we effectively account for the approximate
difference in selection volume for our $z\sim8$ and $z\sim9$ samples.

The simplest way for us to account for any evolution in the $UV$ LF is
through the normalization $\phi^*$ -- since it simply requires that we
compare the number of sources we find in our $z\sim8$ and $z\sim9$
samples with that found in our simulations (see below) to derive the
approximate evolution, i.e.,
\begin{equation}
\phi^* (z=9) = \phi^* (z=8) \frac{n_{obs,z=9}}{n_{obs,z=8}} \frac{n_{no-evol-sim,z=8}}{n_{no-evol-sim,z=9}}
\end{equation}
where $n_{obs,z=9}$ is the number of sources in our $z\sim9$ selection
after correction for contamination (i.e., $\sim$2.3), $n_{obs,z=8}$ is
the number of sources in our $z\sim8$ selection after correction for
contamination (i.e., $\sim$18), $n_{no-evol-sim,z=8}$ is the number of
$z\sim8$ candidates we find in our simulations for our $z\sim8$
selection based on a fiducial lensed LF, and $n_{no-evol-sim,z=9}$ is
the number of $z\sim9$ candidates we find in our simulations for our
$z\sim9$ selection based on this same LF.

As in our previous papers, we estimate the relative numbers of sources
we would expect at both redshifts from simulations.  To perform these
simulations, we insert artificial sources with a variety of redshifts
and luminosities into the real observations and then attempt to select
these objects using our $z\sim8$ and our $z\sim9$-10 selection
criteria.  We generate artificial images for each source in these
simulations using our well-tested cloning software (Bouwens et
al.\ 1998; Bouwens et al.\ 2003; Bouwens et al.\ 2007) which we use to
artificially redshift similar luminosity $z\sim4$ galaxies from the
HUDF to higher redshift.  We scale the size of galaxies at fixed
luminosity as $(1+z)^{-1}$ to match the observed size-redshift scaling
at $z>3$ (e.g., Bouwens et al.\ 2004; Oesch et al.\ 2010; Mosleh et
al.\ 2012).  We take the $UV$-continuum slope $\beta$ of galaxies in
our simulations to have a mean value and $1\sigma$ scatter of $-2.3$
and 0.45, respectively, to match the observed trends extrapolated to
$z\sim8$-9 (Bouwens et al.\ 2012b; Finkelstein et al.\ 2012; Bouwens
et al.\ 2014a).

For simplicity, we estimate the relative numbers of sources we would
expect in both samples without making use of the deflection maps
estimated for all 19 CLASH clusters used in the present search.  As we
demonstrate in Appendix B, we can approximately ignore the impact of
lensing in estimating the selectability of sources, if the quantity we
are interested in calculating is the relative number expected for
sources in two adjacent redshift samples, e.g., $z\sim8$ and
$z\sim9$.  Lensing does not have a big impact on the relative number
of sources seen in two adjacent samples due to the similar impact it
has on the selection efficiencies, volumes, and luminosities of
galaxies in both samples.  Nevertheless, for the simulations we have
run, the relative numbers of lensed $z\sim9$ galaxies to lensed
$z\sim8$ galaxies is slightly lower (16\%) in our simulations than if
we ignore the impact of lensing in calculating its selectability (and
only consider a boost to the LF from some fiducial magnification
factor).

The $UV$ LFs we input into the simulations have the following
parameters: $M_{UV}^{*}=-22.4$, $\alpha=-2.0$, and
$\phi^*=5.5\times10^{-5}$ Mpc$^{-3}$.  These luminosity parameters
were chosen to implicitly include a factor of $\sim$9 magnification
from gravitational lensing -- which is the median magnification
estimated for sources in our selection -- so the effective $M^*$ at
$z\sim8$ is chosen to be $\sim$2.4 mag brighter than seen in blank
field studies (e.g., Oesch et al.\ 2012b).  The faint-end slope we
assume approximately matches what we would expect based on the $UV$ LF
results at $z\sim7$-8 (Bouwens et al.\ 2011b; Oesch et al.\ 2012b;
Bradley et al.\ 2012) which point to faint-end slopes $\alpha$ of
$-2$.  No change is required in the faint-end slope $\alpha$ of the
LF, due to the perfect trade-off between magnification and source
dilution effects for slopes of $-2$ (e.g., Broadhurst et al.\ 1995).
The normalization $\phi^*$ we choose has no effect on our final
results (due to the differential nature of this calculation).  While
the LFs we adopt for these simulations could, in principle, affect our
evolutionary results, the overall size of such effects will be small
due to the differential nature of the comparison we are making.  We
also verified that the surface density of $z\sim8$ sources predicted
by this model LF showed a very similar magnitude dependence as seen
for our $z\sim8$ sample.

Using the above simulation procedure and aforementioned LF, we
repeatedly added artificial sources to the real CLASH observations for
all 19 CLASH clusters, created catalogs, and repeated our $z\sim8$ and
$z\sim9$ selections.  In total, we repeated the described simulations
20 times for each cluster field to obtain an accurate estimate of the
total number of sources (selection volume and redshift distribution)
we would expect to find in each sample, given the described luminosity
function.

\begin{deluxetable*}{ccccc}
\tablewidth{13cm}
\tabletypesize{\footnotesize}
\tablecaption{$UV$ Luminosity Densities and Star Formation Rate Densities to $-17.7$ AB mag (0.05 $L_{z=3} ^{*}$: see \S4.5).\tablenotemark{a,b}\label{tab:sfrdens}}
\tablehead{
\colhead{} & \colhead{} & \colhead{$\textrm{log}_{10} \mathcal{L}$} & \colhead{$\textrm{log}_{10}$ SFR density} \\
\colhead{Dropout} & \colhead{} & \colhead{(ergs s$^{-1}$} & \colhead{($M_{\odot}$ Mpc$^{-3}$ yr$^{-1}$)} \\
\colhead{Sample} & \colhead{$<z>$} & \colhead{Hz$^{-1}$ Mpc$^{-3}$)} & \colhead{Uncorrected} & \colhead{Corrected\tablenotemark{c}}}
\startdata
$J$ & 9.2 & 25.03$_{-0.49}^{+0.37}$ & $-$2.87$_{-0.49}^{+0.37}$ & $-$2.87$_{-0.49}^{+0.37}$ \\
\multicolumn{5}{c}{-------------------------------------------------- }\\
$B$ & 3.8 & 26.42$\pm$0.05 & $-1.48\pm$0.05 & $-1.10\pm0.05$ \\
$V$ & 5.0 & 26.20$\pm$0.06 & $-1.70\pm$0.06 & $-1.36\pm0.06$ \\
$i$ & 5.9 & 25.98$\pm$0.08 & $-1.92\pm$0.08 & $-1.67\pm0.08$ \\
$z$ & 6.8 & 25.84$\pm$0.10 & $-2.06\pm$0.10 & $-1.83\pm0.10$ \\
$Y$ & 8.0 & 25.58$\pm$0.11 & $-2.32\pm$0.11 & $-2.17\pm0.11$ \\
$J$\tablenotemark{d} & 10.0 & 24.45$\pm$0.36 & $-$3.45$\pm$0.36 & $-$3.45$\pm$0.36 \\
\tablenotetext{a}{Integrated down to 0.05 $L_{z=3}^{*}$.  Based upon
  the $z\sim9$ inferred here (Table~\ref{tab:lfparm}: \S4.4) and the
  LF parameters in Oesch et al.\ (2012a,b) and Table 4 of Bouwens et
  al.\ (2014b) (see \S4.5).  The SFR density estimates assume
  $\gtrsim100$ Myr constant SFR and a Salpeter IMF (e.g., Madau et
  al.\ 1998).  Conversion to a Chabrier (2003) IMF would result in a
  factor of $\sim$1.8 (0.25 dex) decrease in the SFR density estimates
  given here.}
\tablenotetext{b}{Uncertainties on the luminosity densities and star
  formation rate densities at $z\sim9$ are calculated by adding in
  quadrature the logarthmic uncertainties on both the $z\sim8$
  densities and the differential evolutionary factors from $z\sim9$
  to $z\sim8$.  Uncertainties on the luminosity densities and star
  formation rate densities at $z\sim4$-8 are computed by marginalizing
  over the likelihood contours for Schechter fits to the $z\sim4$-8
  LFs (from Bouwens et al.\ 2014b).}
\tablenotetext{c}{Dust corrections are from Bouwens et
  al.\ (2014a) and are based on the observed $UV$-continuum slopes.
  No dust correction is assumed at $z\gtrsim9$.}
\tablenotetext{d}{$z\sim10$ determinations and limits are from Bouwens
  et al.\ (2014b).}
\end{deluxetable*}

In total, we find 657 sources that satisfy our $z\sim8$ selection
criteria and 383 sources that satisfy our $z\sim9$ selection criteria,
based on the same luminosity and simulation area (so
$n_{no-evol-sim,z=8}=657$ and $n_{no-evol-sim,z=9}=383$ in Eq. 1
above).  These results imply that without the impact of gravitational
lensing, we would expect to find 58\% as many $z\sim9$ galaxies as
$z\sim8$ galaxies behind our CLASH cluster sample.  While we adopt the
Oesch et al. (2012b) LF in performing this estimate (keeping with our
original treatment), we would have obtained essentially the same
result using other recent $z\sim8$ LF results (e.g. Bouwens et
al.\ 2014; McLure et al.\ 2013; Schenker et al.\ 2013).

As we demonstrate in Appendix B, we would expect to find a very
similar ratio of galaxies in these two samples, even including lensing
in the simulations.  In the case of lensing, we expect just 49\% as
many $z\sim9$ galaxies as $z\sim8$ candidates over our search fields
(though we remark that the precise factor depends slightly [i.e.,
  $\lesssim$20\%] on the lensing model and LF adopted in performing
the calculation).  Therefore, to make a fair comparison between our
$z\sim8$ and $z\sim9$ samples we need to multiply the surface
densities in our $z\sim8$ sample by 0.49 (Eq. 1 above).  In
Figure~\ref{fig:newlf}, we show the comparison of the surface
densities of $z\sim8$ galaxies found over the first 19 CLASH clusters
(corrected for the difference in selection volume) with the surface
densities of $z\sim9$ galaxies found over these clusters.

\subsection{Inferred $UV$ LF at $z\sim9$}

After correction for possible contamination of our selection by
possible low redshift contaminants (see \S3.5), the total number of
$z\sim9$ candidates in our $z\sim9$ selection is 2.3.  This number is
just $0.28_{-0.20}^{+0.39}\times$ the total number of $z\sim8$ sources
to a similar luminosity limit (corrected for differences in the
selection volume: see Figure~\ref{fig:newlf}).  In calculating the
uncertainties on the fraction $0.28_{-0.20}^{+0.39}$, we have
accounted for the Poissonian errors on the total number of galaxies in
the $z\sim8$ and $z\sim9$ samples, as well as the Poissonian
uncertainties in the contamination rates.

Assuming that we can approximate the differences between the $z\sim8$
and $z\sim9$ LFs as occurring simply through density evolution (i.e.,
by changing $\phi^*$), we infer that the value of $\phi^*$ at $z\sim9$
is just $0.28_{-0.20}^{+0.39}\times$ $\phi^*$ at $z\sim8$ (or
$0.28_{-0.26}^{+0.84}\times$ if $2\sigma$ errors are used).  The
present search is inconsistent with no evolution at $>$92\%
confidence.\footnote{Given the approximate degeneracy between
  evolution in $M^*$ and $\phi^*$ for LFs at $z\sim7$-9 where a
  $\Delta M^*=1$ mag change trades off for a $\Delta \phi^*$ change
  (e.g., Figure 8 of Oesch et al.\ 2012b), we could reframe the
  inferred evolution in $\phi^*$ from $z\sim9$ to $z\sim8$ in terms of
  an equivalent evolution in $M^*$ (as we have parameterized the LF
  evolution in the past, e.g., Bouwens et al.\ 2007; Bouwens et
  al.\ 2008; C13).  We estimate that the effective $M^*$ at $z\sim9$
  is $0.5_{-0.3}^{+0.4}$ mag fainter than at $z\sim8$ (keeping
  $\phi^*$ fixed).  However, in a more recent and comprehensive study
  of the $UV$ LFs from $z\sim7$ to $z\sim4$, Bouwens et al.\ (2014b)
  find that the overall evolution can be better represented by an
  evolution in $\phi^*$ and $\alpha$ (with a more limited evolution in
  $M^*$).}

Using the recent Oesch et al.\ (2012b) determination from
HUDF09+CANDELS+ERS field studies (Bouwens et al.\ 2011b; Grogin et
al.\ 2011; Koekemoer et al.\ 2011; Windhorst et al.\ 2011) that
$\phi^*$ at $z\sim8$ is $5.0_{-3.3}^{+7.0}\times10^{-4}$ Mpc$^{-3}$,
we estimate that $\phi^*$ at $z\sim9$ is
$1.4_{-1.1}^{+2.0}\times10^{-4}$ Mpc$^{-3}$.  For the purpose of
parametrizing a $z\sim9$ LF, we will assume that $M^*$ and $\alpha$ at
$z\sim9$ match that derived by Oesch et al.\ (2012b) at $z\sim8$.  The
resultant $z\sim9$ LF is illustrated in the right panel of
Figure~\ref{fig:newlf} and compared with the Oesch et al.\ (2012b)
$z\sim8$ LF.  Use of other recent $z\sim8$ LF results (e.g., Bouwens
et al.\ 2014; Bradley et al.\ 2012; Schenker et al.\ 2013; McLure et
al.\ 2013; Schmidt et al.\ 2014) yield very similar results for $M^*$
($\lesssim$0.3 mag), $\alpha$ ($\lesssim$0.2), and $\phi^*$ ($<0.1$
dex).\footnote{We persist in our reliance of the Oesch et al.\ (2012b)
  LF results to maintain consistency with our original submission (but
  note the overall agreement of the Oesch et al.\ 2012b results with
  more recent determinations of the LF at $z\sim8$).}

What effect will field-to-field variations (i.e., ``cosmic variance'')
have on the overall uncertainties here?  To estimate the size of these
uncertainties, we first considered the case of a single cluster field.
We used the Trenti \& Stiavelli (2008) cosmic variance calculator,
assumed a mean redshift of 8.0 and 9.2 for our two samples (as
estimated from our simulations: see Figure~\ref{fig:zdist}), took the
$\Delta z$ width for these redshift distributions to be 0.8, and
assumed that the relevant area in the source plane was
$0.4'\times0.4'$.  The latter area in the source plane assumes a
factor of $\sim$10 dilution of the total search area (consistent with
the mean magnification factors found here) and further that only 30\%
of the total area on our WFC3/IR images is effective for finding
$z\sim9$ galaxies.  The fractional uncertainty we estimated in our
volume density estimates from field-to-field variations is 0.55 and
0.58 for our $z\sim8$ and $z\sim9$ selections, respectively, over a
single CLASH cluster field.

Since each of our cluster fields provides an independent sightline on
the high redshift universe, we need to reduce the derived variance by
$\sim19^{0.5}\sim4.4$ (though we remark that the actual reduction will
be slightly less than this since all our clusters will not receive
equal weight in the total volume calculation and hence the gains from
independent sightlines will be less).  This results in fractional
uncertainties of $\sim$0.13 in the total number of sources in the
current $z\sim8$ and $z\sim9$ samples.  Since our $z\sim9$ LF estimate
is based on a differential comparison of the present $19$-cluster
$z\sim8$ and $z\sim9$ samples, we must add both of these uncertainties
in quadrature to derive the approximate fractional uncertainty.  The
result is 0.19.  By comparison, $z\sim9$ searches using a single 4.4
arcmin$^2$ deep field would yield a fractional uncertainty of
$\sim$0.5 in the volume density of $z\sim9$ galaxies from large-scale
structure (``cosmic variance'').  This is much higher than the present
uncertainties arising from large-scale structure.

Overall, the uncertainties from large-scale structure only have a
fairly marginal impact on our total uncertainty in $\phi^*$ for the
$z\sim9$ LF, increasing it by just 3\% over what one would estimate
based on the small numbers in the current $z\sim9$ selection.

\subsection{$UV$ Luminosity and Star Formation Rate Density at $z\sim9$}

We can utilize our newly estimated $z\sim9$ LF to determine the
approximate $UV$ luminosity density and SFR density at $z\sim9$-10.
We compute these luminosity densities to a limiting luminosity 0.05
$L_{z=3}^{*}$, which is the effective limit of the Oesch et
al.\ (2012b) $z\sim8$ LF we used as a reference point for inferring
the $z\sim9$ LF.  This limiting luminosity is also what one would
expect for $z\sim8$-9 searches in the CLASH program to $\sim$27 AB mag
assuming a $\sim9\times$ magnification factor -- which is equivalent
to the average magnification factor for $z\sim9$ galaxy candidates
uncovered in the present search.  We can convert the $UV$ luminosities
we estimate to SFR densities using the canonical $UV$
luminosity-to-SFR conversion factor (Madau et al.\ 1998: see also
Kennicutt 1998).

The $z\leq8$ SFR density determinations are corrected for dust
extinction based on the values Bouwens et al.\ (2013b) estimate based
on the observed $UV$-continuum slopes $\beta$.  Given the observed
trends towards bluer $UV$-continuum slopes $\beta$ at very high
redshifts (e.g., Stanway et al.\ 2005; Bouwens et al.\ 2006, 2009,
2012b, 2013b; Finkelstein et al.\ 2012; Wilkins et al.\ 2011), we
would expect the dust extinction at $z\sim9$-10 to be zero, and
therefore apply no dust correction to the SFR density determinations
there.

We present the $UV$ luminosity and SFR densities we estimate at
$z\sim9$ in Figure~\ref{fig:sfrdens} and also in
Table~\ref{tab:sfrdens}.  For context, we also provide the SFR and
$UV$ luminosity densities of several noteworthy determinations in the
literature over the redshift range $z\sim0$ to $z\sim10$ (Bouwens et
al.\ 2007, 2010; Ellis et al.\ 2013; Oesch et al.\ 2013, 2014).  We
also show the SFR density estimates at $z\sim9.6$ and $z\sim10.8$ from
the $z\sim9.6$ Z12 and $z\sim10.8$ C13 candidates, as estimated by
C13.

\subsection{Implications of the present $z\sim9$-10 search for the evolution of the LF at $z>6$}

One of our primary motivations for obtaining constraints on the $UV$
LF at $z\sim9$ was to characterize the evolution of the $UV$ LF at
$z>8$ and to test whether the $UV$ LF at $z>8$ really evolves more
rapidly as a function of redshift -- as recently found by Oesch et
al.\ (2012a: see also Bouwens et al.\ 2012a) -- or the evolution is
more consistent with a simple extrapolation of the $UV$ LF trends
found by Bouwens et al.\ (2011b: see also Bouwens et al.\ 2008) over
the redshift range $z\sim4$-8.  Several theoretical models (Trenti et
al.\ 2010; Lacey et al.\ 2011) support the idea that the $UV$ LF might
indeed evolve faster at $z>8$ as a function of redshift than at
$z\sim4$-8 (e.g., Figure 8 of Oesch et al.\ 2012a or Figure 10 of
Oesch et al.\ 2014), and we want to test this hypothesis using our
current results.

To determine which of these two scenarios the present observations
favor, we first compute the change in $\phi^*$ each would predict.
Using the Bouwens et al.\ (2014b) fitting formula for the evolution of
the $UV$ LF, we estimate an expected change of $\Delta \log_{10}
\phi^* = 0.23$ and $\Delta M_{UV}^* \sim 0.15$ in the $UV$ LF from
$z=8$ to $z=9.2$ (the mean redshift of our sample).  Taking
advantage of the approximate degeneracy between $M^*$ and $\phi^*$ at
$z\sim7$ and $z\sim8$ ($\Delta M^* = 1$ is nearly degenerate with
$\Delta \log_{10} \phi^* = 1$: see Figure 8 of Oesch et al.\ 2012b),
we can convert this to a change in $\phi^*$ over the redshift interval
$z=8$ to $z=9.2$, i.e., $\Delta \log_{10} \phi^* \sim 0.4$ dex
so that $\phi^*(z=9.2) = 0.4 \phi^*(z=8)$.  Oesch et al.\ (2012a) also
estimate the rate of evolution from $z\sim10$ to $z\sim8$ based on
their $z\sim10$ HUDF09+ERS+CANDELS search results, which is more rapid
than implied by the Bouwens et al. (2011b) fitting formula.  The
best-fit evolution in $\phi^*$ that Oesch et al.\ (2012a) find is a
$0.54_{-0.19}^{+0.36}$ dex change per unit redshift, so that
$\phi^*(z=9.2) = 0.23_{-0.15}^{+0.15} \phi^*(z=8)$.

The evolution we measure from $z=9.2$ to $z\sim8$ (\S4.4) is such
that $\phi^*(z=9.2) = 0.28_{-0.20}^{+0.39} \phi^*(z=8)$ (fixed $M^*$
and $\alpha$).  As compared with the two different evolutionary
scenarios, we can see that the observed evolution may suggest
marginally more rapid evolution than seen at lower redshifts
$z\sim4$-8, but is nonetheless consistent that evolution (versus one
would expect utilizing the Bouwens et al.\ 2011b LF fitting formula
where $dM^*/dz\sim0.33$: which is consistent with the new Bouwens et
al.\ 2014b results if one excludes constraints at the bright end from
wide-area searches).

The new ultra-deep WFC3/IR data over the HUDF/XDF field from the
HUDF12 program also tentatively support a more rapid evolution.  Using
a sample of four $z\sim9$ sources, Ellis et al.\ (2013) find
$\phi^*(z=9) = 0.25_{-0.09}^{+0.15} \phi^*(z=8)$ while Oesch et
al.\ (2013) find $\phi^*(z=9) = 0.26_{-0.12}^{+0.15} \phi^*(z=8)$
using similar samples.  Searches to $z\sim10$ (Oesch et al.\ 2013,
2014; Bouwens et al.\ 2014b) are again consistent with a slightly more
rapid evolution.  However, the results are not at all definitive, and
indeed lensed candidate galaxies at redshifts as high as $z\sim10.8$
identified by C13 would appear more consistent with a slower
evolution.  In any case, it seems clear that more observations, such
as available with the Frontier Fields program or further study of the
CANDELS fields (GO 13792: PI Bouwens) will be required to resolve this
situation.

\section{Summary}

In this paper, we have explored the use of a two-color Lyman-Break
selection to search for $z\sim9$-10 galaxies in the first 19 clusters
observed with the CLASH program.  Building on the important
exploratory studies of Z12 and C13, we extend the CLASH $z\sim9$-10
selections even deeper to the approximate magnitude limit of the CLASH
program ($\sim$27 mag).  Such a search is possible making full use of
the noteworthy Spitzer/IRAC observations over the CLASH clusters
(Egami et al.\ 2008; Bouwens et al.\ 2011c), allowing us to determine
which $z\sim9$-10 galaxy candidates have a blue spectral slope redward
of the break (and therefore strongly favor a $z\sim9$-10 solution) and
which candidates do not.

In total, we find three plausible $z\sim9$-10 galaxy candidates from
the CLASH program that satisfy a two-color Lyman-Break-like selection
criteria (i.e., $(J_{110}+J_{125})/2-H_{160}>0.7$ and
$JH_{140}-H_{160}<0.5$) and have a combined $JH_{140}+H_{160}$ S/N of
$\geq$6.0.  The $H_{160,AB}$ magnitudes for sources in our selection
range from $\sim$25.7 AB mag to 26.9 AB mag.  The candidates are found
behind the galaxy clusters MACSJ1149.6+2223, MACSJ1115.9+0129, and
MACSJ1720.3+3536.  The highest S/N source in our selection is the
$z\sim9.6$ Z12 candidate (here $z_{ph}\sim9.7$).  All three of our
candidates have reasonably blue $H_{160,AB}-IRAC$ colors strongly
favoring the $z\sim9$-10 solution for all three sources in our
selection.

As in other $z\sim9$ and $z\sim10$ selections we have performed
(Bouwens et al.\ 2011a; Z12; C13; Oesch et al.\ 2013, 2014; Bouwens et
al.\ 2014b), we have carefully considered the possibility of
contamination.  We find that the only significant source of
contamination is from the ``photometric scatter'' of lower redshift
galaxies into our selection and that this likely contributes only
$\sim$0.7 source to our $z\sim9$-10 sample (\S3.5), with MACS1720-JD1
or MACS1115-JD1 being the most probable contaminant.  However, we
emphasize that we cannot completely exclude the possibility that the
contamination rate may be somewhat higher.

To determine the implications of the present search results for the
$UV$ LF, $UV$ luminosity density, and SFR density at $z\sim9$-10, we
introduce a novel differential approach for deriving these quantities.
Our procedure is to simply compare the number of candidate $z\sim9$
galaxies found in the CLASH fields with the number of $z\sim8$
galaxies found in the CLASH fields and then correct this ratio for the
relative selection volume at $z\sim8$ and $z\sim9$.  This procedure
takes advantage of the fact that the ratio of selection volumes at
$z\sim8$ and $z\sim9$ for a given cluster is not greatly dependent on
details of the gravitational lensing model one is utilizing (e.g., see
Figure~\ref{fig:relvol}).  This procedure therefore provides us with a
very robust technique for measuring the evolution of the $UV$ LF to
$z>9$ using searches over lensing cluster fields.  The $z\sim8$ and
$z\sim9$ selection volumes we derive are from detailed simulations
where artificial sources are added to the real imaging data and then
reselected using the same criteria as applied to the real data
(\S4.3).

Comparing our sample of three candidate $z\sim9$ galaxies with a
sample of 19 $z\sim8$ galaxies found to similar $6\sigma$ detection
significance over the same CLASH cluster fields (and correcting for
the expected 23\% contamination in our $z\sim9$ selection), we derive
the approximate evolution in the $UV$ LF to $z\sim9$.  One strength of
the present evolutionary estimate is that we are particularly
insensitive to large-scale structure uncertainties due to our many
independent lines of sight on the high redshift universe (\S4.4).

We find that $\phi^*$ for the $z\sim9$ LF is just
$0.28_{-0.20}^{+0.39}\times$ the equivalent $\phi^*$ at $z\sim8$
(\S4.4: keeping $M^*$ and $\alpha$ fixed).  We would have expected the
normalization $\phi^*$ of the LF at $z\sim9$ to be just $0.4\times$
that at $z\sim8$, if the evolution in the $UV$ LF proceeded at the
same rate as seen at $z\sim4$-8.  While the present result is
consistent with there being no significant change in the rate of
evolution of the $UV$ LF from $z\sim9$ to $z\sim4$, our result does
favor slightly more rapid evolution of the LF at $z>8$, as suggested
by Oesch et al.\ (2012a) based on their early search results for
$z\sim10$ galaxies.  Using the best-fit evolutionary trend from Oesch
et al.\ (2012a: see also Oesch et al.\ 2014), we would have predicted
the normalization of our $z\sim9$ LF to be
$0.23_{-0.15}^{+0.15}\times$ that at $z\sim8$.  Several theoretical
models (Trenti et al.\ 2010; Lacey et al.\ 2011) support the idea that
the $UV$ LF may evolve faster at $z>8$ as a function of redshift than
at $z\sim4$-8 (see Figure 8 of Oesch et al.\ 2012a and Figure 10 of
Oesch et al.\ 2014).  Despite the excellent agreement between the
present evolutionary result and new findings from Oesch et
al.\ (2012a), Ellis et al.\ (2013), and Oesch et al.\ (2014), the
uncertainties on the evolution of the LF at $z>8$ are still somewhat
large.

In the future, we expect further advances in our constraints on the
$UV$ LF at $z\geq9$ from the Frontier Fields program (Lotz et
al.\ 2014)\footnote{http://www.stsci.edu/hst/campaigns/frontier-fields/}$^{,}$\footnote{http://www.stsci.edu/hst/campaigns/frontier-fields/HDFI \_SWGReport2012.pdf},
pointing follow-up observations of $z\sim9$-10 candidates over CANDELS
(GO 13792: PI Bouwens), and the new wide-area BoRG program (GO 13767:
PI Trenti).  Substantially deeper Spitzer observations over the CLASH
clusters, as part of the Surf's Up program (Bradac et al.\ 2012) and
other programs, should allow us both to obtain better constraints on
the nature of current $z\sim9$ candidates and to provide initial
estimates of the stellar mass density at $z\sim9$.

\acknowledgements

We are grateful for extensive feedback on our manuscript from Pascal
Oesch and Garth Illingworth.  We thank Anton Koekemoer for providing
us with high quality reductions of the available HST observations of
our CLASH cluster fields.  We acknowledge Dan Magee for assisting with
the reductions of the IRAC data for our clusters using MOPEX.  We are
greatly appreciative to Ryan Quadri for sending us his results for the
rest-frame V band LF of red $z\sim2$ galaxies based on a UKIDSS UDS
search.  Comments by the anonymous referee significantly improved this
paper.  We acknowledge support from ERC grant HIGHZ \#227749, an NWO
vrij competitie grant, and the NASA grant for the CLASH MCT program.
The dark cosmology center is funded by the DNRF.

\appendix

\section{A.  Description of The Redshift Priors}

In \S3.4, we present redshift likelihood distributions for the three
candidate $z\sim9$ galaxies in our selection.  This allows us to
estimate the relative probability that sources in this sample
correspond to higher or lower redshift galaxies
(Figure~\ref{fig:seds}).  However, in doing so, we must utilize a
prior.  We consider three different redshift priors: (1) a flat
redshift-independent prior, (2) a prior calibrated to published LF or
LF trends, and (3) a prior tuned to reproduce the results from our
photometric scattering experiments (\S3.5).  This section discusses
the latter two priors in detail.

\textit{LF-calibrated Prior:} The second prior we consider is
calibrated according to published LFs or LF trends.  For this prior,
we give special attention to two galaxy populations: star-forming
galaxies at $z\sim9$ and faint red galaxies at $z\sim1.3$-2.  These
are the only two galaxy populations which can at least provide
approximate fits to the sources in our selection and therefore have
nominal $\chi^2$'s that are not especially large.  For the faint red
$z\sim1.3$-2 galaxy case, we calibrate our priors based on the LF
results of Giallongo et al.\ (2005) for red galaxies using deep
near-IR observations available over the HDF-North and HDF-South fields
(Williams et al.\ 1996; Casertano et al.\ 2000) and the K20
spectroscopic sample (Cimatti et al.\ 2002).  At $z\sim2$, their
$<m^*/m(bimodal)$ LF results correspond to $M_{B,0}^{*}=-21.90$ mag,
$\phi^*=2\times10^{-4}$ Mpc$^{-3}$ mag$^{-1}$, and $\alpha=-0.53$.
The basic validity of these LF results has been verified with much
improved statistics based on new results for red galaxies over the
UKIDSS Ultra Deep Survey field (Lawrence et al.\ 2007) where fits
yield $M_{V}=-21.9$, $\phi^*=2\times10^{-4}$ Mpc$^{-3}$, and
$\alpha=0.07$ (R. Quadri et al.\ 2012, private communication).
Meanwhile, at $z\sim1.3$, the Giallongo et al.\ (2005)
$<m^*/m(bimodal)$ LF results correspond to $M_{B,0}^{*}=-21.49$ mag,
$\phi^*=5\times10^{-4}$ Mpc$^{-3}$ mag$^{-1}$, and $\alpha=-0.53$.
Finally, for the $z\sim9$ star-forming galaxy case, our priors use the
Bouwens et al. (2011b) LF fitting formula as a guide (which is a
parameterization of the evolution of the $UV$ LF from $z\sim8$ to
$z\sim4$: see \S7.5 of that paper).

Assuming a deep blank search at $\sim$28.5 mag (the approximate
intrinsic magnitude of our candidates after correction for
magnification) with a $\Delta z\sim1$, $\Delta \textrm{mag} \sim 1$
selection window, we find that these LFs predict $\sim$1.2 $z\sim9$
galaxies per arcmin$^{2}$, but 0.14 faint red galaxies per
arcmin$^{2}$ over the redshift range $z\sim1.3$-2.5.  Surprisingly
enough, these results suggest that we would be much more likely (i.e.,
by $\sim$9$\times$) to find a blue galaxy at $z\sim9$ with our
selection than a faint red galaxy at $z\sim1.3$-2.  Even correcting
these predictions based on the present search results for $z\sim9$
galaxies (where we find just $\sim$55$_{-38}^{+75}$\% as many galaxies
as expected from the Bouwens et al.\ 2011b fitting formula), $z\sim9$
galaxies would still be $5\times$ more abundant on the sky at
$\sim28.5$ mag than red (old and/or dusty) galaxies at $z\sim1.3$-2.5.
For the purposes of our ``LF calibrated'' prior, we will assume that
$z\sim9$ galaxies have a $5\times$ higher surface density on the sky
than $z\sim1.3$-2.5 red (old and/or dusty) galaxies.

\textit{Contamination-Tuned Prior:} Of course, it is not simply the
faint red (old and/or dusty) galaxies at $z\sim1.3$-2 that can
contaminate $z>8$ selections.  Other galaxies can scatter into
$z\sim9$ selections through noise.  This makes the low-redshift
solution more likely than what we would calculate based on
observationally-based LFs.  Considered by themselves, each
photometrically-scattered source would be unlikely to look very much
like a probable $z\sim9$ candidate, but one must account for the fact
that there are some $\sim4\times10^{4}$ sources in our fields which
noise could conspire to make look like such a $z\sim9$-10 candidate.
We account for this possibility with our third ``contamination tuned''
prior.  With this prior, we adjust the relative likelihood of the high
and low redshift peaks for our entire three source $z\sim9$-10 galaxy
sample so that it matches the 23\% contamination rate estimated in our
photometric scattering experiments described in
\S3.5.\footnote{Admittedly, a more accurate approach would be to
  determine the actual redshift distribution of the
  intermediate-magnitude sources scattering into our selection and
  present it in Figure~\ref{fig:seds}.} However, it is worth keeping
in mind that results based on the third prior likely overweight the
probability that sources are low-redshift contaminants.  This is due
to our photometric scatter simulations not accounting for the fact
that red (old and/or dusty) galaxies are even rarer at $\sim$27-28 mag
than the $\sim$24-25.5 magnitude sources we use as inputs to our
photometric scattering simulations (e.g., Figure 11 from Oesch et
al.\ 2012a).

\section{B.  Estimating the ratio of effective volumes for different selections behind lensing clusters using similar volumes in blank field searches}

In deriving the differential evolution in the $UV$ LF from $z\sim9$
to $z\sim8$, one particularly significant assumption we made in \S4.2
was that the relative numbers of galaxies expected to be present in
our $z\sim9$ and $z\sim8$ samples would remain roughly the same
whether or not we included lensing in the simulations.

In this section, we test the accuracy of this assumption by making use
of four different gravitational lensing models and galaxy clusters
from the CLASH program (i.e., MACS1149, MACS1115, MACS1720, and
MACS0416) when creating mock galaxy fields.  We then create mock
galaxy fields over the same cluster ignoring the lensing deflection
fields.  By comparing the relative number of $z\sim8$ and $z\sim9$
galaxies we select from the two simulations, we test the assumption we
made in \S4.2 of this paper.

In simulating the mock galaxy fields subject to lensing, we use
exactly the same set of assumptions that we used for the simulations
described in \S4.2.  Starting with the same (and non-evolving) LF of
galaxies at $z\sim8$ and $z\sim9$, we construct a mock catalog of
sources over multiple cluster fields.  We then create artificial
images of each sources by artificially redshifting similar luminosity
$z\sim4$ galaxies from the HUDF to higher redshift.  While redshifting
the sources, we scale their sizes as $(1+z)^{-1}$ at fixed luminosity
and take the $UV$-continuum slope $\beta$ of galaxies in our
simulations to have a mean value and $1\sigma$ scatter of $-2.3$ and
0.45, respectively, to match the observed trends extrapolated to
$z\sim8$-9.  We remap the simulated images of galaxies in the source
plane onto the image plane using the lensing models we have available
for these clusters (Zitrin et al.\ 2012).  We then added these
simulated fields to the actual CLASH observations and attempted to
recover the mock sources using the same $z\sim8$ and $z\sim9$
selection criteria as given in this paper.

\begin{figure}
\epsscale{0.6} \plotone{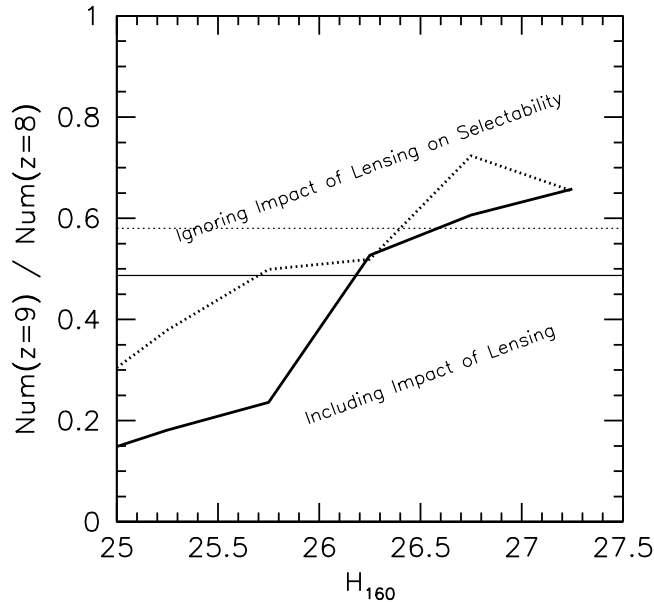}
\caption{The relative number of galaxies we expect to be present in
  our $z\sim9$ and $z\sim8$ samples assuming no evolution including
  the impact of lensing on surface brightness profiles and
  selectability of individual galaxies in simulations behind galaxy
  clusters (\textit{thick solid line}) and ignoring the impact of
  lensing in the simulations (\textit{thick dashed line}), as a
  function of $H_{160}$-band magnitude (see Appendix B).  The thin
  lines show the average relative numbers averaging over the full
  magnitude range.  In the simulations where we ignore lensing, a
  factor-of-9 magnification boost to the LF is nonetheless considered
  (see \S4.2).  The slight magnitude dependence to the relative
  numbers is due to slight differences in the distance modulus to the
  two samples and due to the slightly smaller sizes of $z\sim9$
  galaxies in our simulations.  In both the case where the effects of
  lensing are fully included in estimating the selectability of
  individual sources and where this is ignored, we expect
  approximately the same ratio of galaxies in the two
  samples.\label{fig:wwolensing}}
\end{figure}

In Figure~\ref{fig:wwolensing}, we present the relative surface
density of $z\sim8$ and $z\sim9$ galaxies we recover from
non-evolving LF from our CLASH observations and compare with the
surface density of galaxies we find without including lensing (but
assuming a uniform factor-of-9 magnification in the luminosity of all
sources: see \S4.2).  Overall, we find that the ratio of galaxies we
select in the two cases is similar, but not exactly the same.

In fact, the relative numbers of $z\sim9$ galaxies to $z\sim8$
galaxies is slightly ($\sim$16\%) lower in simulations where we
include the effect of lensing.  We find similar results for all four
cluster lensing models we have run simulations, but note that the
relative numbers can show a dependence on the LFs or average
magnification factors we assume ($\sim$10-20\%).  However, for all
reasonable LF or magnification factors we consider, the relative
numbers do not differ by $\gtrsim$20\% from the ratios we give here.

\section{C.  $z\sim8$ Comparison Sample}

In deriving the $UV$ LF at $z\sim9$ from our CLASH search results
(\S4), we make use of a baseline sample of $z\sim8$ candidate galaxies
in CLASH that we contrast with a similar $z\sim9$ sample to establish
the evolution from $z\sim8$ to $z\sim9$.

We tabulate this sample of $z\sim8$ candidate galaxies in
Table~\ref{tab:samplez8}.  Four of the 18 candidates from our
selection are reported as $z\sim8$ candidates by Bradley et
al.\ (2014), five candidates are reported as $z\sim7$ candidates by
Bradley et al.\ (2014), four candidates do not appear in the Bradley
et al.\ (2014) $z\sim6$-8 compilation, and three candidates are found
over CLASH clusters not considered by Bradley et al.\ (2014).

\begin{deluxetable}{ccc}
\tablewidth{0cm}
\tabletypesize{\footnotesize}
\tablecaption{Candidate $z\sim8$ galaxies selected from CLASH to
  compare with a similar selection at $z\sim9$\label{tab:samplez8}}
\tablehead{\colhead{Right Ascension} & \colhead{Declination} &
  \colhead{$H_{160}$}} \startdata
17:22:25.76 & 32:08:58.2 & 26.7$\pm$0.2\tablenotemark{a,d}\\
13:47:30.47 & $-$11:45:27.6 & 26.5$\pm$0.2\tablenotemark{b}\\
13:47:33.89 & $-$11:45:09.3 & 26.1$\pm$0.2\tablenotemark{a}\\
13:47:31.82 & $-$11:44:13.20 & 25.4$\pm$0.1\\
21:29:24.92 & $-$07:42:04.2 & 26.8$\pm$0.2\\
03:29:40.34 & $-$02:12:44.9 & 26.3$\pm$0.2\\
06:47:59.14 & 70:14:15.2 & 26.4$\pm$0.1\tablenotemark{b}\\
06:47:40.49 & 70:14:15.3 & 26.5$\pm$0.2\\
06:47:40.90 & 70:14:41.5 & 25.8$\pm$0.1\tablenotemark{a}\\
06:47:40.67 & 70:14:56.5 & 26.0$\pm$0.2\tablenotemark{b}\\
06:47:44.76 & 70:15:37.9 & 27.1$\pm$0.2\tablenotemark{b}\\
11:15:52.85 & 01:28:56.7 & 25.5$\pm$0.1\\
11:15:51.09 & 01:30:34.9 & 25.5$\pm$0.1\tablenotemark{a}\\
15:32:58.81 & 30:21:02.2 & 25.3$\pm$0.1\tablenotemark{b}\\
19:31:45.22 & $-$26:34:24.6 & 25.9$\pm$0.2\tablenotemark{d}\\
21:29:40.29 & 00:06:15.2 & 26.7$\pm$0.2\tablenotemark{c}\\
01:31:48.72 & $-$13:36:49.1 & 26.3$\pm$0.2\tablenotemark{c}\\
01:31:54.51 & $-$13:36:00.5 & 25.4$\pm$0.2\tablenotemark{c}\\
\enddata
\tablenotetext{a}{Present in the $z\sim8$ sample of Bradley et al.\ (2014).}
\tablenotetext{b}{Present in the $z\sim7$ sample of Bradley et al.\ (2014).}
\tablenotetext{c}{Search field not considered in Bradley et al.\ (2014).}
\tablenotetext{d}{This $z\sim8$ candidate is sufficiently compact
  (i.e., the SExtractor stellarity parameter is $>$0.9 in at least 2
  of the 5 near-IR bands probed [where 1 and 0 corresponds to a point
    source and extended source, respectively]) that it may correspond
  to a low-mass star.}
\end{deluxetable}

\begin{thebibliography}{} 
\bibitem[Anders 
\& Fritze-v.~Alvensleben(2003)]{2003A&A...401.1063A} Anders, P., \& Fritze-v.~Alvensleben, U.\ 2003, \aap, 401, 1063
\bibitem[Arnouts et al.(1999)]{1999MNRAS.310..540A} Arnouts, S., Cristiani, 
S., Moscardini, L., et al.\ 1999, \mnras, 310, 540
\bibitem[Atek et al.(2010)]{2010ApJ...723..104A} Atek, H., Malkan, M., 
McCarthy, P., et al.\ 2010, \apj, 723, 104
\bibitem[Atek et al.(2011)]{2011ApJ...743..121A} Atek, H., Siana, B., 
Scarlata, C., et al.\ 2011, \apj, 743, 121
\bibitem[Ben{\'{\i}}tez(2000)]{2000ApJ...536..571B} Ben{\'{\i}}tez, N.\ 
2000, \apj, 536, 571
\bibitem[Bertin and Arnouts (1996)]{1996A&AS..117..393B} Bertin, E.\ and 
Arnouts, S.\ 1996, \aaps, 117, 39
\bibitem[Blanton \& Roweis(2007)]{2007AJ....133..734B} Blanton, M.~R.,
  \& Roweis, S.\ 2007, \aj, 133, 734
\bibitem[Boone et al.(2011)]{2011A&A...534A.124B} Boone, F., Schaerer,
  D., Pell{\'o}, R., et al.\ 2011, \aap, 534, A124
\bibitem[Bouwens, Broadhurst and Silk (1998)]{1998ApJ...506..557B} Bouwens,
R., Broadhurst, T.\ and Silk, J.\ 1998, \apj, 506, 557
\bibitem[Bouwens et al.(2003)]{2003ApJ...593..640B} Bouwens, R., 
Broadhurst, T., \& Illingworth, G.\ 2003, \apj, 593, 640 
\bibitem[Bouwens et al.(2003)]{2003ApJ...595..589B} Bouwens, R.~J., 
Illingworth, G.~D., Rosati, P., et al.\ 2003, \apj, 595, 589
\bibitem[Bouwens et al.(2004)]{2004ApJ...611L...1B} Bouwens, R.~J., 
Illingworth, G.~D., Blakeslee, J.~P., Broadhurst, T.~J., 
\& Franx, M.\ 2004, \apjl, 611, L1
\bibitem[Bouwens et al.(2004)]{2004ApJ...616L..79B} Bouwens, R.~J., 
Thompson, R.~I., Illingworth, G.~D., et al.\ 2004, \apjl, 616, L79 
\bibitem[Bouwens et al. (2006)]{2006Bouwens} Bouwens, R.J., Illingworth,
G.D., Blakeslee, J.P., \& Franx, M.  2006a, \apj, 653, 53 
\bibitem[Bouwens \& Illingworth(2006)]{2006Natur.443..189B} Bouwens,
  R.~J., \& Illingworth, G.~D.\ 2006b, \nat, 443, 189
\bibitem[Bouwens et al.(2007)]{2007ApJ...670..928B} Bouwens, R.~J., 
Illingworth, G.~D., Franx, M., \& Ford, H.\ 2007, \apj, 670, 928
\bibitem[Bouwens et al.\ (2008)]{2008ApJ...686..230B} Bouwens, R.~J., 
Illingworth, G.~D., Franx, M., \& Ford, H.\ 2008, \apj, 686, 230 
\bibitem[Bouwens et al.\ (2009)]{2009ApJ...705..936B} Bouwens, R.~J., et
  al.\ 2009, \apj, 705, 936
\bibitem[Bouwens et al.(2010)]{2010ApJ...709L.133B} Bouwens, R.~J., 
Illingworth, G.~D., Oesch, P.~A., et al.\ 2010, \apjl, 709, L133
\bibitem[Bouwens et al.(2011)]{2011Natur.469..504B} Bouwens, R.~J., 
Illingworth, G.~D., Labb{\'e}, I., et al.\ 2011a, \nat, 469, 504 
\bibitem[Bouwens et al.(2011)]{2011ApJ...737...90B} Bouwens, R.~J., 
Illingworth, G.~D., Oesch, P.~A., et al.\ 2011b, \apj, 737, 90 
\bibitem[Bouwens et al.(2011)]{2011sptz.prop80168B} Bouwens, R., Zheng, W., 
Moustakas, L., et al.\ 2011c, Spitzer Proposal, 80168 
\bibitem[Bouwens et al.(2012)]{2012ApJ...752L...5B} Bouwens, R.~J., 
Illingworth, G.~D., Oesch, P.~A., et al.\ 2012a, \apjl, 752, L5 
\bibitem[Bouwens et al.(2012)]{2012ApJ...754...83B} Bouwens, R.~J., 
Illingworth, G.~D., Oesch, P.~A., et al.\ 2012b, \apj, 754, 83
\bibitem[Bouwens et al.(2013)]{2013ApJ...765L..16B} Bouwens, R.~J., Oesch, 
P.~A., Illingworth, G.~D., et al.\ 2013, \apjl, 765, L16
\bibitem[Bouwens et al.(2014a)]{2013arXiv1306.2950B} Bouwens, R.~J., 
Illingworth, G.~D., Oesch, P.~A., et al.\ 2014a, \apj, in press, arXiv:1306.2950 
\bibitem[Bouwens et al.(2014)]{2014arXiv1403.4295B} Bouwens, R.~J., 
Illingworth, G.~D., Oesch, P.~A., et al.\ 2014b, \apj, submitted, arXiv:1403.4295 
\bibitem[Bowler et al.(2012)]{2012MNRAS.426.2772B} Bowler, R.~A.~A., 
Dunlop, J.~S., McLure, R.~J., et al.\ 2012, \mnras, 426, 2772
\bibitem[Brada{\v c} et al.(2012)]{2012sptz.prop90009B} Brada{\v c},
  M., Gonzalez, A., Schrabback, T., et al.\ 2012, Spitzer Proposal,
  90009
\bibitem[Brada{\v c} et al.(2014)]{2014ApJ...785..108B} Brada{\v c}, M., 
Ryan, R., Casertano, S., et al.\ 2014, \apj, 785, 108 
\bibitem[Bradley et al.(2012)]{2012arXiv1204.3641B} Bradley, L.~D., Trenti, 
M., Oesch, P.~A., et al.\ 2012, \apj, 760, 108
\bibitem[Bradley et al.(2014)]{2014ApJ...792...76B} Bradley, L.~D., Zitrin, 
A., Coe, D., et al.\ 2014, \apj, 792, 76
\bibitem[Brammer et al.(2008)]{2008ApJ...686.1503B} Brammer, G.~B., van 
Dokkum, P.~G., \& Coppi, P.\ 2008, \apj, 686, 1503
\bibitem[Brammer et al.(2013)]{2013ApJ...765L...2B} Brammer, G.~B., van 
Dokkum, P.~G., Illingworth, G.~D., et al.\ 2013, \apjl, 765, L2 
\bibitem[Broadhurst et al.(1995)]{1995ApJ...438...49B} Broadhurst, T.~J., 
Taylor, A.~N., \& Peacock, J.~A.\ 1995, \apj, 438, 49 
\bibitem[Bruzual 
\& Charlot(2003)]{2003MNRAS.344.1000B} Bruzual, G., \& Charlot, S.\ 2003, \mnras, 344, 1000 
\bibitem[Bunker et al.(2003)]{2003MNRAS.342L..47B} Bunker, A.~J.,
  Stanway, E.~R., Ellis, R.~S., McMahon, R.~G., \& McCarthy,
  P.~J.\ 2003, \mnras, 342, L47
\bibitem[Bunker et al.(2010)]{2010MNRAS.409..855B} Bunker, A.~J., Wilkins, 
S., Ellis, R.~S., et al.\ 2010, \mnras, 409, 855
\bibitem[Capak et al.(2013)]{2013ApJ...773L..14C} Capak, P., Faisst, A., 
Vieira, J.~D., et al.\ 2013, \apjl, 773, L14
\bibitem[Casertano et al.(2000)]{2000AJ....120.2747C} Casertano, S., de 
Mello, D., Dickinson, M., et al.\ 2000, \aj, 120, 2747
\bibitem[Chabrier(2003)]{2003PASP..115..763C} Chabrier, G.\ 2003, \pasp, 
115, 763 
\bibitem[Cimatti et al.(2002)]{2002A&A...381L..68C} Cimatti, A.,
  Daddi, E., Mignoli, M., et al.\ 2002, \aap, 381, L68
\bibitem[Coe et al.(2006)]{2006AJ....132..926C} Coe, D., Ben{\'{\i}}tez, 
N., S{\'a}nchez, S.~F., et al.\ 2006, \aj, 132, 926
\bibitem[Coe et al.(2013)]{2013ApJ...762...32C} Coe, D., Zitrin, A., 
Carrasco, M., et al.\ 2013, \apj, 762, 32 (C13)
\bibitem[Coleman et al.(1980)]{1980ApJS...43..393C} Coleman, G.~D., Wu, 
C.-C., \& Weedman, D.~W.\ 1980, \apjs, 43, 393
\bibitem[Cushing et al.(2005)]{cushing2005} Cushing, M.C., Rayner,
  J.T., \& Vacca, W.D.  2005, \apj, 623, 1115
\bibitem[Dickinson et al.(2000)]{2000ApJ...531..624D} Dickinson, M., 
Hanley, C., Elston, R., et al.\ 2000, \apj, 531, 624
\bibitem[Dickinson et al.(2004)]{2004ApJ...600L..99D} Dickinson, M., Stern, 
D., Giavalisco, M., et al.\ 2004, \apjl, 600, L99 
\bibitem[Dow-Hygelund et al.(2007)]{2007ApJ...660...47D} Dow-Hygelund, 
C.~C., et al.\ 2007, \apj, 660, 478
\bibitem[Egami et al.(2008)]{2008sptz.prop60034E} Egami, E., Ellis, R., 
Fazio, G., et al.\ 2008, Spitzer Proposal, 60034 
\bibitem[Ellis et al.(2013)]{2013ApJ...763L...7E} Ellis, R.~S., McLure, 
R.~J., Dunlop, J.~S., et al.\ 2013, \apjl, 763, L7
\bibitem[Fazio et al.(2004)]{2004ApJS..154...10F} Fazio, G.~G., Hora, 
J.~L., Allen, L.~E., et al.\ 2004, \apjs, 154, 10 
\bibitem[Ferguson et al.(2004)]{2004ApJ...600L.107F} Ferguson, H.~C.~et 
al.\ 2004, \apjl, 600, L107
\bibitem[Finkelstein et al.(2012)]{2012ApJ...756..164F} Finkelstein, S.~L., 
Papovich, C., Salmon, B., et al.\ 2012, \apj, 756, 164 
\bibitem[Finkelstein et al.(2013)]{2013Natur.502..524F} Finkelstein, S.~L., 
Papovich, C., Dickinson, M., et al.\ 2013, \nat, 502, 524
\bibitem[Fioc 
\& Rocca-Volmerange(1997)]{1997A&A...326..950F} Fioc, M., \& Rocca-Volmerange, B.\ 1997, \aap, 326, 950 
\bibitem[Fontana et al.(2010)]{2010ApJ...725L.205F} Fontana, A., Vanzella, 
E., Pentericci, L., et al.\ 2010, \apjl, 725, L205
\bibitem[Giallongo et al.(2005)]{2005ApJ...622..116G} Giallongo, E., 
Salimbeni, S., Menci, N., et al.\ 2005, \apj, 622, 116
\bibitem[Giavalisco et al.(2004)]{2004ApJ...600L.103G} Giavalisco, M., 
Dickinson, M., Ferguson, H.~C., et al.\ 2004, \apjl, 600, L103
\bibitem[Grazian et 
al.(2006)]{2006A&A...449..951G} Grazian, A., Fontana, A., de Santis, C., et al.\ 2006, \aap, 449, 951
\bibitem[Grogin et al.(2011)]{2011ApJS..197...35G} Grogin, N.~A., Kocevski, 
D.~D., Faber, S.~M., et al.\ 2011, \apjs, 197, 35 
\bibitem[Hayes et al.(2012)]{2012MNRAS.425L..19H} Hayes, M., Laporte, N., 
Pell{\'o}, R., Schaerer, D., \& Le Borgne, J.-F.\ 2012, \mnras, 425, L19 
\bibitem[Ilbert et 
al.(2006)]{2006A&A...457..841I} Ilbert, O., Arnouts, S., McCracken, H.~J., et al.\ 2006, \aap, 457, 841 
\bibitem[Ilbert et al.(2009)]{2009ApJ...690.1236I} Ilbert, O., Capak, P., 
Salvato, M., et al.\ 2009, \apj, 690, 1236
\bibitem[Iye et al.(2006)]{2006Natur.443..186I} Iye, M., Ota, K., 
Kashikawa, N., et al.\ 2006, \nat, 443, 186
\bibitem[Kennicutt(1998)]{1998ARA&A..36..189K} Kennicutt, R.~C., Jr.\ 1998, \araa, 36, 189 
\bibitem[Kirkpatrick et al.(2012)]{2012ApJ...753..156K} Kirkpatrick, J.~D., 
Gelino, C.~R., Cushing, M.~C., et al.\ 2012, \apj, 753, 156
\bibitem[Koekemoer et al.(2003)]{2003hstc.conf..337K} Koekemoer, A.~M., 
Fruchter, A.~S., Hook, R.~N., 
\& Hack, W.\ 2003, HST Calibration Workshop : Hubble after the Installation of the ACS and the NICMOS Cooling System, 337 
\bibitem[Koekemoer et al.(2011)]{2011ApJS..197...36K} Koekemoer, A.~M., 
Faber, S.~M., Ferguson, H.~C., et al.\ 2011, \apjs, 197, 36 
\bibitem[Kotulla et al.(2009)]{2009MNRAS.396..462K} Kotulla, R., Fritze, 
U., Weilbacher, P., \& Anders, P.\ 2009, \mnras, 396, 462
\bibitem[Kron (1980)]{kron} Kron, R. G. 1980, \apjs, 43, 305
\bibitem[Labb{\'e} et al.(2006)]{2006ApJ...649L..67L} Labb{\'e}, I., 
Bouwens, R., Illingworth, G.~D., \& Franx, M.\ 2006, \apjl, 649, L67
\bibitem[Labb{\'e} et al.(2010)]{2010ApJ...708L..26L} Labb{\'e}, I., et 
al.\ 2010a, \apjl, 708, L26
\bibitem[Labb{\'e} et al.(2010)]{2009arXiv0911.1356L} Labb{\'e}, I., et
  al.\ 2010b, \apjl, 716, L103
\bibitem[Labb{\'e} et al.(2013)]{2013ApJ...777L..19L} Labb{\'e}, I., Oesch, 
P.~A., Bouwens, R.~J., et al.\ 2013, \apjl, 777, L19
\bibitem[Lacey et al.(2011)]{2011MNRAS.412.1828L} Lacey, C.~G., Baugh, 
C.~M., Frenk, C.~S., \& Benson, A.~J.\ 2011, \mnras, 412, 1828 
\bibitem[Laidler et al.(2007)]{2007PASP..119.1325L} Laidler, V.~G., 
Papovich, C., Grogin, N.~A., et al.\ 2007, \pasp, 119, 1325 
\bibitem[Laporte et al.(2011)]{2011A&A...531A..74L} Laporte, N.,
  Pell{\'o}, R., Schaerer, D., et al.\ 2011, \aap, 531, A74
\bibitem[Laporte et al.(2012)]{2012A&A...542L..31L} Laporte, N.,
  Pell{\'o}, R., Hayes, M., et al.\ 2012, \aap, 542, L31
\bibitem[Laporte et al.(2014)]{2014A&A...562L...8L} Laporte, N.,
  Streblyanska, A., Clement, B., et al.\ 2014, \aap, 562, L8
\bibitem[Lawrence et al.(2007)]{2007MNRAS.379.1599L} Lawrence, A., Warren, 
S.~J., Almaini, O., et al.\ 2007, \mnras, 379, 1599 
\bibitem[Lorenzoni et al.(2011)]{2011MNRAS.414.1455L} Lorenzoni, S., 
Bunker, A.~J., Wilkins, S.~M., et al.\ 2011, \mnras, 414, 1455
\bibitem[Lotz et al.(2014)]{2014AAS...22325401L} Lotz, J., Mountain, M., 
Grogin, N.~A., et al.\ 2014, American Astronomical Society Meeting 
Abstracts, 223, \#254.01
\bibitem[Madau et al.(1996)]{1996MNRAS.283.1388M} Madau, P., Ferguson, 
H.~C., Dickinson, M.~E., et al.\ 1996, \mnras, 283, 1388 
\bibitem[Madau et al.\ (1998)]{mad98} Madau, P., Pozzetti, L. \&
Dickinson, M. 1998, \apj, 498, 106
\bibitem[Maizy et 
al.(2010)]{2010A&A...509A.105M} Maizy, A., Richard, J., de Leo, M.~A., Pell{\'o}, R., \& Kneib, J.~P.\ 2010, \aap, 509, A105 
\bibitem[Makovoz 
\& Khan(2005)]{2005ASPC..347...81M} Makovoz, D., \& Khan, I.\ 2005, Astronomical Data Analysis Software and Systems XIV, 347, 81 
\bibitem[McLure et al.(2010)]{2010MNRAS.403..960M} McLure, R.~J., Dunlop, 
J.~S., Cirasuolo, M., et al.\ 2010, \mnras, 403, 960
\bibitem[McLure et al.(2013)]{2013MNRAS.432.2696M} McLure, R.~J., Dunlop, 
J.~S., Bowler, R.~A.~A., et al.\ 2013, \mnras, 432, 2696
\bibitem[Meurer et al.(1999)]{1999ApJ...521...64M} Meurer, G.~R., Heckman, 
T.~M., \& Calzetti, D.\ 1999, \apj, 521, 64 
\bibitem[Mosleh et al.(2012)]{2012ApJ...756L..12M} Mosleh, M., Williams, 
R.~J., Franx, M., et al.\ 2012, \apjl, 756, L12 
\bibitem[Narayan 
\& Bartelmann(1996)]{1996astro.ph..6001N} Narayan, R., \& Bartelmann, M.\ 1996, arXiv:astro-ph/9606001
\bibitem[Oesch et al.(2010)]{2010ApJ...709L..21O} Oesch, P.~A., Bouwens, 
R.~J., Carollo, C.~M., et al.\ 2010, \apjl, 709, L21
\bibitem[Oesch et al.(2012)]{2012ApJ...745..110O} Oesch, P.~A., Bouwens, 
R.~J., Illingworth, G.~D., et al.\ 2012a, \apj, 745, 110 
\bibitem[Oesch et al.(2012)]{2012ApJ...759..135O} Oesch, P.~A., Bouwens, 
R.~J., Illingworth, G.~D., et al.\ 2012b, \apj, 759, 135
\bibitem[Oesch et al.(2013)]{2013ApJ...773...75O} Oesch, P.~A., Bouwens, 
R.~J., Illingworth, G.~D., et al.\ 2013, \apj, 773, 75
\bibitem[Oesch et al.(2014)]{2014ApJ...786...108O} Oesch, P.~A., Bouwens, 
R.~J., Illingworth, G.~D., et al.\ 2014, \apj, 786, 108
\bibitem[Oke \& Gunn(1983)]{1983ApJ...266..713O} Oke, J.~B., \& Gunn, 
J.~E.\ 1983, \apj, 266, 713 
\bibitem[Ono et al.(2012)]{2012ApJ...744...83O} Ono, Y., Ouchi, M., 
Mobasher, B., et al.\ 2012, \apj, 744, 83
\bibitem[Polletta et al.(2007)]{2007ApJ...663...81P} Polletta, M., Tajer, 
M., Maraschi, L., et al.\ 2007, \apj, 663, 81 
\bibitem[Popesso et al.(2009)]{2009A&A...494..443P} Popesso, P., et al.\ 2009, \aap, 494, 443 
\bibitem[Postman et al.(2012)]{2012ApJS..199...25P} Postman, M., Coe, D., 
Ben{\'{\i}}tez, N., et al.\ 2012, \apjs, 199, 25
\bibitem[Rayner et al.(2009)]{2009Rayner} Rayner, J.T., Cushing, M.C.,
  \& Vacca, W.D.  2009, \apjs, 185, 289
\bibitem[Reddy \& Steidel(2009)]{2009ApJ...692..778R} Reddy, N.~A., \&
  Steidel, C.~C.\ 2009, \apj, 692, 778
\bibitem[Richard et al.(2008)]{2008ApJ...685..705R} Richard, J., Stark, 
D.~P., Ellis, R.~S., et al.\ 2008, \apj, 685, 705
\bibitem[Santos et al.(2004)]{2004ApJ...606..683S} Santos, M.~R., Ellis, 
R.~S., Kneib, J.-P., Richard, J., \& Kuijken, K.\ 2004, \apj, 606, 683 
\bibitem[Schechter(1976)]{1976ApJ...203..297S} Schechter, P.\ 1976, \apj, 
203, 297
\bibitem[Schenker et al.(2012)]{2012ApJ...744..179S} Schenker, M.~A., 
Stark, D.~P., Ellis, R.~S., et al.\ 2012, \apj, 744, 179
\bibitem[Schenker et al.(2013)]{2013ApJ...768..196S} Schenker, M.~A., 
Robertson, B.~E., Ellis, R.~S., et al.\ 2013, \apj, 768, 196 
\bibitem[Schiminovich et al.(2005)]{2005ApJ...619L..47S} Schiminovich, D.,
et al.\ 2005, \apjl, 619, L47 
\bibitem[Scoville et al.(2007)]{2007ApJS..172....1S} Scoville, N., Aussel, 
H., Brusa, M., et al.\ 2007, \apjs, 172, 1
\bibitem[Shapley et al.(2005)]{2005ApJ...626..698S} Shapley, A.~E., 
Steidel, C.~C., Erb, D.~K., et al.\ 2005, \apj, 626, 698 
\bibitem[Silva et al.(1998)]{1998ApJ...509..103S} Silva, L., Granato, 
G.~L., Bressan, A., \& Danese, L.\ 1998, \apj, 509, 103
\bibitem[Sirianni et al.(2005)]{2005PASP..117.1049S} Sirianni, M., et al.\ 
2005, \pasp, 117, 1049 
\bibitem[Stanway et al.(2003)]{2003MNRAS.342..439S} Stanway, E.~R., Bunker, 
A.~J., \& McMahon, R.~G.\ 2003, \mnras, 342, 439 
\bibitem[Stanway et al.(2005)]{2005MNRAS.359.1184S} Stanway, E.~R., 
McMahon, R.~G., \& Bunker, A.~J.\ 2005, \mnras, 359, 1184
\bibitem[Stark et al.(2010)]{2010MNRAS.408.1628S} Stark, D.~P., Ellis, 
R.~S., Chiu, K., Ouchi, M., \& Bunker, A.\ 2010, \mnras, 408, 1628 
\bibitem[Steidel et al.(1996)]{1996ApJ...462L..17S} Steidel, C.~C., 
Giavalisco, M., Pettini, M., Dickinson, M., 
\& Adelberger, K.~L.\ 1996, \apjl, 462, L17 
\bibitem[Steidel et al.\ (1999)]{1999ApJ...519....1S} Steidel, C.\ C.,
Adelberger, K.\ L., Giavalisco, M., Dickinson, M.\ and Pettini, M.\ 1999,
\apj, 519, 1
\bibitem[Steidel et al.(2003)]{2003ApJ...592..728S} Steidel, C.~C., 
Adelberger, K.~L., Shapley, A.~E., Pettini, M., Dickinson, M., 
\& Giavalisco, M.\ 2003, \apj, 592, 728
\bibitem[Stutz et al.(2008)]{2008ApJ...677..828S} Stutz, A.~M., Papovich, 
C., \& Eisenstein, D.~J.\ 2008, \apj, 677, 828
\bibitem[Szalay et al.(1999)]{1999AJ....117...68S} Szalay, A.~S.,
Connolly, A.~J., \& Szokoly, G.~P.\ 1999, \aj, 117, 68
\bibitem[Trenti 
\& Stiavelli(2008)]{2008ApJ...676..767T} Trenti, M., \& Stiavelli, M.\ 2008, \apj, 676, 767 
\bibitem[Trenti et al.(2010)]{2010ApJ...714L.202T} Trenti, M., Stiavelli, 
M., Bouwens, R.~J., et al.\ 2010, \apjl, 714, L202
\bibitem[Trenti et al.(2011)]{2011ApJ...727L..39T} Trenti, M., Bradley, 
L.~D., Stiavelli, M., et al.\ 2011, \apjl, 727, L39
\bibitem[van der Wel et al.(2011)]{2011ApJ...742..111V} van der Wel, A., 
Straughn, A.~N., Rix, H.-W., et al.\ 2011, \apj, 742, 111 
\bibitem[Vanzella et al.(2009)]{2009ApJ...695.1163V} Vanzella, E., et al.\ 
2009, \apj, 695, 1163 
\bibitem[Whitelock et al.(1995)]{1995MNRAS.276..219W} Whitelock, P., 
Menzies, J., Feast, M., et al.\ 1995, \mnras, 276, 219 
\bibitem[Wilkins et al.(2011)]{2011MNRAS.417..717W} Wilkins, S.~M., Bunker, 
A.~J., Stanway, E., Lorenzoni, S., \& Caruana, J.\ 2011, \mnras, 417, 717 
\bibitem[Williams et al.(1996)]{1996AJ....112.1335W} Williams, R.~E., 
Blacker, B., Dickinson, M., et al.\ 1996, \aj, 112, 1335
\bibitem[Windhorst et al.(2011)]{2011ApJS..193...27W} Windhorst, R.~A., 
Cohen, S.~H., Hathi, N.~P., et al.\ 2011, \apjs, 193, 27 
\bibitem[Wuyts et al.(2008)]{2008ApJ...682..985W} Wuyts, S., Labb{\'e}, I., 
Schreiber, N.~M.~F., et al.\ 2008, \apj, 682, 985 
\bibitem[Yan \& Windhorst(2004)]{2004ApJ...612L..93Y} Yan, H., \&
  Windhorst, R.~A.\ 2004, \apjl, 612, L93
\bibitem[Yan et al.(2010)]{2010RAA....10..867Y} Yan, H.-J., Windhorst, 
R.~A., Hathi, N.~P., et al.\ 2010, Research in Astronomy and Astrophysics, 
10, 867
\bibitem[Zheng et al.(2012)]{2012Natur.489..406Z} Zheng, W., Postman, M., 
Zitrin, A., et al.\ 2012, \nat, 489, 406 (Z12) 
\bibitem[Zitrin et al.(2014)]{2014arXiv1407.3769Z} Zitrin, A., Zheng, W., 
Broadhurst, T., et al.\ 2014, \apj, in press, arXiv:1407.3769
\end{thebibliography}
\end{document}